\documentclass[aps,prd,onecolumn,preprintnumbers,groupedaddress,showpacs,nofootinbib,amssymb]{revtex4-2}
\usepackage{graphicx,color}
\usepackage{amsmath}
\usepackage{amssymb}
\usepackage{amsfonts}
\usepackage{bm}
\usepackage{cancel}
\usepackage{slashed}
\usepackage{hyperref}

\usepackage{tikz}

\newcommand{\e}{\mathrm{e}}

\allowdisplaybreaks[4]
\begin{document}

\preprint{KEK-TH-2835, KEK-Cosmo-0421}

\title{Observational signatures of negative mass wormholes through their shadows}

\author{Shin'ichi~Nojiri$^{1,2}$}
\email{nojiri@nagoya-u.jp}
\affiliation{$^{1)}$ KEK Theory Center, Institute of Particle and Nuclear Studies,
High Energy Accelerator Research Organization (KEK), Oho 1-1, Tsukuba, Ibaraki 305-0801, Japan}
\affiliation{$^{1)}$ Kobayashi-Maskawa Institute for the Origin of Particles and the Universe, Nagoya University, Nagoya 464-8602, Japan}

\author{Sergei~D.~Odintsov$^{3,4}$}
\email{odintsov@ice.csic.es} 
\affiliation{$^{3)}$  Institute of Space Sciences, (ICE-CSIC), Carrer de Can Magrans s/n, 08193 Barcelona, Spain}
\affiliation{$^{4)}$ Instituci\'o Catalana de Recerca i Estudis Avan\c{c}ats (ICREA),
Passeig Luis Companys, 23, 08010 Barcelona, Spain}

\author{Diego~S\'aez-Chill\'on~G\'omez$^{5,6}$}
\email{diego.saez@uva.es}
\affiliation{$^{5)}$ Department of Theoretical, Atomic and Optical
Physics, and Laboratory for Disruptive Interdisciplinary Science (LaDIS), Campus Miguel Delibes, \\ 
University of Valladolid UVA, Paseo Bel\'en, 7, 47011
Valladolid, Spain}
\affiliation{$^{6)}$ Department of Physics, Universidade Federal do Cear\'a (UFC), Campus do Pici, Fortaleza - CE, C.P. 6030, 60455-760 - Brazil}

\begin{abstract}
Systems containing objects with negative mass (NMOs) are considered. Such a system consists of one object with positive mass and one NMO, where a bound state exists even though the force exerted by the NMO on the object with positive-mass is repulsive. 
Unlike standard binaries composed of positive-mass objects, the emitted gravitational waves exhibit decreasing frequency and amplitude over time.
We propose a model that removes the ghost appearing in the construction of the Ellis–Bronnikov wormhole, a candidate for a NMO.
Furthermore, we perform numerical simulations to obtain the optical appearance of such NMOs. The observed luminosity is also compared with the Schwarzschild black hole and with the Simpson-Visser wormhole, revealing clear differences in the photon ring substructure around the central object. 
\end{abstract}

\maketitle

\section{Introduction}

Although the negative mass object (NMO) is not always consistent with physics, such an object has been studied for a long time. 
Luttinger's essay ~\cite{Luttinger1951B} might be the first research article in which the term ``negative mass'' appeared. 
After that, the dynamics of the system, which includes NMOs, was studied in detail by Bondi~\cite{Bondi:1957zz}, and extensive research has been carried out 
\cite{Farnes:2017gbf, Belletete:2013nqa, Bonnor:1989, Cebeci:2006hx, Mann:1997jb, Forward:1990, Moller:2017yzv, Khamehchi:2017, Socas-Navarro:2019pps, Petit:2014ura, Najera:2021tcx, Manfredi:2018nlx, Trivedi:2026knt}. 
In~\cite{Ellis:1973yv, Bronnikov:1973fh}, a wormhole solution was presented, which behaves as a positive mass object in one universe and as an NMO in another universe. 
The two universes are connected by a wormhole throat. 
See also~\cite{Huang:2020qmn}. 
The lensing by NMOs was investigated in \cite{Cramer:1994qj, Safonova:2001nd, Safonova:2001vz, Eiroa:2001zz, Safonova:2002si, Shaikh:2017zfl}, and the abundance of NMOs was also studied in \cite{Takahashi:2013jqa}. 
Furthermore, zero mass objects were considered in \cite{Chetouani:1984qdm, Perlick:2003vg, Abe:2010ap, Nakajima:2012pu, Toki:2011zu, Tsukamoto:2012xs, Tsukamoto:2013dna, Yoo:2013cia, Izumi:2013tya, Nakajima:2014nba, Bozza:2015haa, Bozza:2015wbw, Tsukamoto:2017hva, Tsukamoto:2017hva, Bozza:2017dkv, Asada:2017vxl}. 
More recently, some observational tests on gravitational waves from NMO binaries were proposed in Ref.~\cite{Trivedi:2026knt}. 

The existence of NMOs is generally regarded as being inconsistent with conventional physics. 
As well known from classical mechanics of the elementary course, the Lagrangian of a particle located at $\bm{r}$ with mass $m$ and the potential $V\left(\bm{r}\right)$ is given by 
\begin{align}
\label{clsclprtcl}
L=\frac{1}{2} m \dot{\bm{r}}\cdot \dot{\bm{r}} - V\left(\bm{r}\right) \, .
\end{align}
In the case that the mass $m$ is negative, a ghost appears in the corresponding quantum mechanics rendering the model inconsistent. 
Recently, it has been argued that NMOs may consistently arise \cite{Nojiri:2026ubx}, whereas Ref.~\cite{Trivedi:2026knt} investigated possible observational signatures of negative-mass binaries.  
A familiar analogy is provided by an air bubble in water, where the bubble looks as if it has a negative mass because the water around the bubble falls due to gravity. 
Of course, no ghost appears in the system of the bubble and water. 
Furthermore, the existence of NMOs might not be so unnatural if gravity is taken into account. 
We should note that the negative cosmological constant is identified with negative energy density. 
When there is another cosmological fluid besides the negative cosmological constant, we may consider the situation where the total energy density is almost equal to zero. 
A fluctuation in the fluid may make the density of the fluid smaller in some regions. 
Then the regions behave as if there are NMOs, like the bubble in the water. 
Although such fluctuations would not generally be stable, if we consider the situation where there is a usual positive point mass in the fluid with negative pressure, regions where the energy density is negative may appear.
This is because the pressure gradient generated by the negative pressure balances the gravitational attraction generated by the positive point mass. 
And as a result, the point mass pushes away the fluid. 
Such a configuration is generically stable. 
In \cite{Nojiri:2026ubx}, the spacetime with an NMO was constructed without ghosts by using two scalar fields. 

Moreover, the released images by the Event Horizon Telescope (EHT) of the supermassive black holes at the centre of the M87 and Milky Way galaxies has stimulated research on gravitational physics~\cite{EventHorizonTelescope:2019dse, EventHorizonTelescope:2022wkp}, making it possible to study black holes directly from the images. 
The reconstructed pictures show a central shadow surrounded by a complex structure of light rings, which is the consequence of strong magnetic fields and the emission of the plasma that composes the accretion disk~\cite{Chan:2014nsa} but also due to the high deflection that light experiences near a black hole~\cite{Perlick:2003vg, Gralla:2019xty, Perlick:2021aok}. 
Despite all available evidence indicating that both the central objects in M87 and the Milky Way are actually Kerr black holes, the possibility of observing other non-Kerr objects in the future remains open. 
Then, motivated by this possibility, a large number of analyses have been conducted to test the Kerr paradigm by simulating the observed luminosity of an accretion disk~\cite{Guerrero:2021ues, Younsi:2021dxe, Vagnozzi:2022moj}. 
Such luminosity is strongly affected, neglecting accretion disk properties, by the structure of the spacetime, such as the locations of circular orbits for photons, their number and the shape of the potential for null geodesics~\cite{Olmo:2021piq, Olmo:2025ctf}. Moreover, the possibility of getting some evidence of extra dimensions through the images have been analyzed \cite{Banerjee:2019nnj,Banerjee:2022jog}. 

In this paper, by starting with the tutorial review of the classical mechanics of the system including NMOs, we find that there exists a bound system composed of one positive mass object and one NMO, despite the repulsive force exerted by the NMO on the positive-mass object. 
Although this system emits gravitational waves, its behaviour differs significantly from that of ordinary binaries, implying a decreasing frequency and amplitude of the gravitational wave with time. 
There is a candidate for an NMO, that is, the Ellis-Boronnikov wormhole~\cite{Ellis:1973yv, Bronnikov:1973fh}, which was proposed over forty years ago. 
The wormhole was constructed by using a non-canonical scalar field, which is a ghost. The rotating case of this wormhole has been also studied in Ref.~\cite{Chew:2016epf}.
We propose a model in which the ghost is eliminated. 
After reviewing the lensing effect by the NMOs,  by using ray-tracing techniques, we show the numerical results for the observed image of NMOs or the Ellis-Bronnikov wormhole.

In the next Section, we review the mechanics of the system, including NMOs, at a tutorial level. 
Especially, we show that the system of one positive mass object and one NMO can make a bound system, although the positive mass object suffers a repulsive force from the NMO. 
In Section~\ref{GW}, we investigate the gravitational wave emitted by the bound system and find that the frequency and the amplitude of the gravitational wave decrease in time, which is different from the behaviour of the standard gravitational wave emitted by the system of two positive mass objects. 
The Ellis-Boronnikov wormhole is a candidate for an NMO, but it has been constructed by using a non-canonical scalar field, which is a ghost. 
In Section~\ref{EBwormhole}, we show that the ghost can be eliminated by using a constraint. 
In Section~\ref{Phtn}, we review the lensing effect by the NMOs, and we observe that the NMOs play the role of the concave lens as expected. 
In Section~\ref{Shadows}, by using the so-called ray-tracing technique, we numerically investigate the observed image of NMOs or the Ellis-Bronnikov wormhole. 
The last section is devoted to the summary and discussion. 

\section{Undergraduate mechanics of negative mass object(s)}\label{udrgrdtmch}

Let us analyze a system through undergraduate mechanics. 
Here, as an approximation, we use Newton's law of mechanics and Newton's law of gravity. 
We should note that the kinetic energy of the NMO is unbounded from below and consequently when the absolute value of the kinetic energy becomes large, the particle picture turns out invalid. 

\subsection{General setup}

As in introductory lectures of classical mechanics, we start with the general setup of a system composed of two particles. 
However, in a system that includes NMO(s), the total mass and the reduced mass can be negative.  We are assuming that there are two point masses with position vectors $\bm{r}_1$ and $\bm{r}_2$ and masses $m_1$ and $m_2$. 
Then, the equations of motion are given by 
\begin{align}
\label{eom}
m_1 \ddot{\bm{r}}_1 = - \frac{G m_1 m_2 \left( \bm{r}_1 - \bm{r}_2 \right)}{\left| \bm{r}_1 - \bm{r}_2 \right|^3}\, , \quad
m_2 \ddot{\bm{r}}_2 = \frac{G m_1 m_2 \left( \bm{r}_1 - \bm{r}_2 \right)}{\left| \bm{r}_1 - \bm{r}_2 \right|^3}\, ,
\end{align}
which lead to the first integral, 
\begin{align}
\label{mmtm}
0= m_1 \ddot{\bm{r}}_1 + m_2 \ddot{\bm{r}}_2
= \frac{d}{dt} \left( m_1 \dot{\bm{r}}_1 + m_2 \dot{\bm{r}}_2 \right) \, , 
\end{align}
that shows the conservation of the total momentum, as usual. 
We now define the total mass $M$, the reduced mass $\mu$, the position vector of the center of mass $\bm{R}$ and the relative position vector $\bm{r}$ as follows, 
\begin{align}
\label{MmuRr}
M\equiv m_1 + m_2 \, , \quad 
\mu \equiv \frac{m_1 m_2}{m_1 + m_2} \, , \quad 
\bm{R} \equiv \frac{m_1 \bm{r}_1 + m_2 \bm{r}_2}{m_1 + m_2} \, , \quad 
\bm{r} \equiv \bm{r}_2 - \bm{r}_1 \, .
\end{align}
Then, we find 
\begin{align}
\label{eom2}
M \ddot{\bm{R}} = 0 \, , \quad 
\mu \ddot{\bm{r}} = - \frac{G \mu M \bm{r}}{\left|\bm{r}\right|^3}\, . 
\end{align}
We may also define the total energy $E$ by 
\begin{align}
\label{tE}
E\equiv&\, \frac{1}{2} m_1 \left| \dot{\bm{r}}_1 \right|^2 + \frac{1}{2} m_2 \left| \dot{\bm{r}}_2 \right|^2 - \frac{Gm_1 m_2}{\left| \bm{r_1} - \bm{r_2} \right|} 
= E_\mathrm{COM} + E_\mathrm{rel}\, , \nonumber \\
&\, E_\mathrm{COM} \equiv \frac{1}{2} M \left| \dot{\bm{R}} \right|^2 \, , \quad 
E_\mathrm{rel} \equiv \frac{1}{2} \mu \left| \dot{\bm{r}} \right|^2 - \frac{G\mu M}{\left|\bm{r}\right|} \, .
\end{align}
Note $E$, $E_\mathrm{COM}$ and $E_\mathrm{rel}$ are conserved, respectively. 

The second equation in \eqref{eom2} tells that the motion of the particle(s) is generated by the central force with respect to the relative position vector $\bm{r}$. 
Therefore, the angular momentum $\bm{L}$ with respect to the relative position vector $\bm{r}$, 
\begin{align}
\label{angmm}
\bm L = \mu \bm{r} \times \dot{\bm{r}}\, ,  
\end{align}
is conserved. 
Then the motion is restricted to a plane perpendicular to $\bm{L}$,which can be chosen to be $xy$-plane, where the following polar coordinates are assumed, 
\begin{align}
\label{plrcrdnt}
x=r\cos\theta\, , \quad y=r\sin\theta \, .
\end{align}
Then we find 
\begin{align}
\label{angmm2}
\bm L = \left(0,0, l\right) = \left(0,0, \mu r^2\dot\theta \right)\, .
\end{align}
Here $l$ is a constant that corresponds to the absolute value of the angular momentum $\bm{L}$, which is conserved. 
For a while, we assume $l>0$. 

The last expression in Eq.~\eqref{tE} has the following form, 
\begin{align}
\label{tE2}
E_\mathrm{rel} = \frac{1}{2} \mu \left( {\dot r}^2 + r^2 {\dot{\theta}}^2 \right) - \frac{G\mu M}{r} \, .
\end{align}
Furthermore by using \eqref{angmm2}, we may remove $\dot\theta$ in \eqref{tE2}, 
\begin{align}
\label{tE3}
E_\mathrm{rel} = \frac{1}{2} \mu {\dot r}^2 + U(r)\, , \quad 
U(r) \equiv \frac{l^2}{2\mu r^2} - \frac{G\mu M}{r} 
= \frac{l^2}{2\mu} \left( \frac{1}{r} - \frac{G\mu^2 M}{l^2} \right)^2 - \frac{G^2\mu^3 M^2}{2l^2} \, .
\end{align}
The expression is an analogue of the particle moving in a one-dimensional line with a potential $U(r)$.
We further rewrite \eqref{tE3} as follows, 
\begin{align}
\label{tE4}
&\, \frac{2E_\mathrm{rel}}{\mu} - \frac{l^2}{\mu^2} \left( \frac{1}{r} - \frac{G\mu^2 M}{l^2} \right)^2 + \frac{G^2\mu^2 M^2}{l^2} \nonumber \\
=&\, \frac{l^2}{\mu^2} \left( \frac{1}{r} - \frac{G\mu^2 M}{l^2} - \sqrt{\frac{\mu^2}{l^2}\left(\frac{G^2\mu^2M^2}{l^2} + \frac{2E_\mathrm{rel}}{\mu}\right)} \right)  
\left( \frac{1}{r} - \frac{G\mu^2 M}{l^2} + \sqrt{\frac{\mu^2}{l^2}\left(\frac{G^2\mu^2M^2}{l^2} + \frac{2E_\mathrm{rel}}{\mu}\right)} \right) 
= {\dot r}^2 \geq 0\, .
\end{align}
Eq.~\eqref{tE4} tells 
\begin{align}
\label{tE5}
\frac{2E_\mathrm{rel}}{\mu} + \frac{G^2\mu^2 M^2}{l^2} \geq 0 \, .
\end{align}
When $\frac{2E_\mathrm{rel}}{\mu} = - \frac{G^2\mu^2 M^2}{l^2} < 0$, the orbit becomes a circle with a radius given by 
\begin{align}
\label{crcl}
r = \frac{l^2}{G\mu^2 M}\, .
\end{align}
When $\frac{2E_\mathrm{rel}}{\mu}>0$, the orbit is restricted to 
\begin{align}
\label{hyp}
\frac{1}{r}> \frac{G\mu^2 M}{l^2} + \sqrt{\frac{\mu^2}{l^2}\left(\frac{G^2\mu^2M^2}{l^2} + \frac{2E_\mathrm{rel}}{\mu}\right)}\, ,
\end{align}
which corresponds to a hyperbolic trajectory. 
When $\frac{2E_\mathrm{rel}}{\mu}=0$, the orbit is a parabolic. 
When $0> \frac{2E_\mathrm{rel}}{\mu} > - \frac{G^2\mu^2 M^2}{l^2}$, the orbit is restricted to 
\begin{align}
\label{elp}
\frac{G\mu^2 M}{l^2} + \sqrt{\frac{\mu^2}{l^2}\left(\frac{G^2\mu^2M^2}{l^2} + \frac{2E_\mathrm{rel}}{\mu}\right)} > 
\frac{1}{r}> \frac{G\mu^2 M}{l^2} - \sqrt{\frac{\mu^2}{l^2}\left(\frac{G^2\mu^2M^2}{l^2} + \frac{2E_\mathrm{rel}}{\mu}\right)}\, ,
\end{align}
which corresponds to an ellipse.  

The shape of the orbit can be found by using \eqref{angmm2} and rewriting Eq.~\eqref{tE3} as follows
\begin{align}
\label{tE6}
E_\mathrm{rel} = \frac{l^2}{2\mu r^4}\left(\frac{dr}{d\theta} \right)^2 + U(r) \, .
\end{align}
By solving \eqref{tE6} with respect to $r=r(\theta)$, we obtain 
\begin{align}
\label{trjctry}
r= \frac{\frac{l^2}{G\mu^2 M}}{1 + \cos\left(\theta - \theta_0\right) \sqrt{ 1 + \frac{2l^2 E_\mathrm{rel}}{G^2 \mu^3 M^2}}}\, .
\end{align}
Here $\theta_0$ is a constant of integration. 
Therefore when $\frac{E_\mathrm{rel}}{\mu}>0$, the orbit becomes a hyperbola, 
when $\frac{E_\mathrm{rel}}{\mu}=0$, the orbit becomes a parabola, 
when $0> \frac{E_\mathrm{rel}}{\mu} > - \frac{G^2\mu^2 M^2}{2l^2}$, the orbit becomes an ellipse 
and when $\frac{E_\mathrm{rel}}{\mu} = - \frac{G^2\mu^2 M^2}{2l^2}<0$, the orbit becomes a circle. 

We may consider the case $l=0$, where \eqref{tE2} has the following form, 
\begin{align}
\label{tE2l0}
E_\mathrm{rel} = \frac{1}{2} \mu {\dot r}^2 - \frac{G\mu M}{r} \, .
\end{align}
This requires 
\begin{align}
\label{tE2l0B}
\frac{E_\mathrm{rel}}{\mu} + \frac{GM}{r} \geq 0 \, .
\end{align}
When $l=0$, the two objects move on the straight line between them and the objects might collide with each other. 

When $\frac{E_\mathrm{rel}}{\mu M}>0$, Eq.~\eqref{tE2l0B} implies $M>0$. 
Then, Eq.~\eqref{tE2l0} can be solved as follows, 
\begin{align}
\label{solpls}
t - t_0 = \mp \sqrt{ \frac{\mu}{2E_\mathrm{rel}}}\frac{G\mu M}{2 E_\mathrm{rel}} \left\{ - \frac{2\sqrt{1 + \frac{G\mu M}{E_\mathrm{rel} r}}}{\frac{G\mu M}{E_\mathrm{rel} r}} 
+ \ln \frac{\sqrt{1 + \frac{G\mu M}{E_\mathrm{rel} r}} + 1}{\sqrt{1 + \frac{G\mu M}{E_\mathrm{rel} r}} -1} \right\}\, ,
\end{align}
Here, $t_0$ is a constant of integration.  When $r\to 0$, we obtain 
\begin{align}
\label{pr0}
t - t_0 \to 0 \, .
\end{align}
Then, two objects of masses $m_1$ and $m_2$ might collide with each other in a finite time, but both objects might move in opposite directions.

When $\frac{E_\mathrm{rel}}{\mu M} <0$, there are the case where $M$ is positive and the case with negative $M$. 
For the case $M>0$, we obtain, 
\begin{align}
\label{solngtv2}
t - t_0 = \mp \sqrt{ - \frac{\mu}{2E_\mathrm{rel}}}\frac{G\mu M}{2 E_\mathrm{rel}} \left\{ \frac{2\sqrt{- 1 - \frac{G\mu M}{E_\mathrm{rel} r}}}{\frac{G\mu M}{E_\mathrm{rel} r}} 
+ 2 \arctan \frac{1}{\sqrt{-1 - \frac{G\mu M}{E_\mathrm{rel} r}}} \right\} \, ,
\end{align}
Whereas for $r\to 0$, it yields
\begin{align}
\label{nr0}
t - t_0 \to 0 \, .
\end{align}
Two objects with masses $m_1$ and $m_2$ can collide with each other in a finite time again.   We also note that Eq.~\eqref{tE2l0B} tells that $r$ has an upper limit, 
\begin{align}
\label{rmx}
r \leq r_\mathrm{max} \equiv - \frac{G\mu M}{E_\mathrm{rel}} \, .
\end{align}
Therefore, the system is bounded and the two objects must collide with each other. 

On the other hand when $\frac{E_\mathrm{rel}}{\mu M} <0$ and $M<0$, we obtain, 
\begin{align}
\label{solngtv}
t - t_0 = \mp \sqrt{ - \frac{\mu}{2E_\mathrm{rel}}}\frac{G\mu M}{2 E_\mathrm{rel}} \left\{ - \frac{2\sqrt{1 + \frac{G\mu M}{E_\mathrm{rel} r}}}{\frac{G\mu M}{E_\mathrm{rel} r}} 
+ \ln \frac{\sqrt{1 + \frac{G\mu M}{E_\mathrm{rel} r}} + 1}{\sqrt{1 + \frac{G\mu M}{E_\mathrm{rel} r}} - 1} \right\} \, ,
\end{align}
Eq.~\eqref{tE2l0} tells that $r$ has a minimum given by
\begin{align}
\label{nmmnm}
r \geq r_\mathrm{min} \equiv - \frac{G\mu M}{E_\mathrm{rel}} \, .
\end{align}
Then, for this case, two objects of masses $m_1$ and $m_2$ cannot collide with each other and do not form a bound system. 

For completeness, we may consider the case of $E_\mathrm{rel}=0$ in \eqref{tE2l0}.  For consistency, one finds $M\geq 0$.  In addition, as $r$ is a constant for $M=0$, we may further assume $M>0$. Then, the solution of \eqref{tE2l0} is given by 
\begin{align}
\label{tE2l0E0sl}
r^\frac{3}{2} = \frac{3}{2} \sqrt{ 2GM} \left( t - t_0 \right) \, . 
\end{align}
Here, $t_0$ is a constant of integration again.  We might set $r=0$ for $t=t_0$, and therefore the two objects might collide with each other in a finite time. 

\subsection{$m_1$, $m_2>0$ case}

First, we consider the case where both $m_1$ and $m_2$ are positive, as any classical mechanics system.  In this case, $M$ and $\mu$ are also positive. 
Therefore, if $l>0$,  when $E_\mathrm{rel}>0$, the orbit becomes a hyperbola,  when $E_\mathrm{rel}=0$, the orbit becomes a parabola, 
when $0> E_\mathrm{rel} > - \frac{G^2\mu^3 M^2}{2l^2}$, the orbit becomes an ellipse  and when $E_\mathrm{rel} = - \frac{G^2\mu^3 M^2}{2l^2}<0$, the orbit becomes a circle. 
Even if $l=0$, $E_\mathrm{rel}\geq 0$, the system is not bounded but it becomes bound when $E_\mathrm{rel}<0$, as in \eqref{rmx}. 
In the case of $l=0$, the two objects might collide with each other. 

\subsection{$m_1$, $m_2<0$ case}

When both mases, $m_1$ and $m_2$, are negative, $M$ and $\mu$ are also negative. 
Eq.~\eqref{tE2} implies that $\frac{E_\mathrm{rel}}{\mu}$ is always positive. 
Therefore, the only possible orbit when $l>0$ is a hyperbola, and any bound state does not arise.  As $\frac{E_\mathrm{rel}}{\mu M}<0$, for $l=0$, the solution is given by \eqref{solngtv}. 
There is a minimum for $r$ as in \eqref{nmmnm}. 

\subsection{$m_1>0$, $m_2<0$ and $m_1 + m_2 \neq 0$ case}

When $m_1>0$, $m_2<0$ and $m_1 + m_2 \neq 0$, $\mu$ and $M$ can be positive or negative. 

First, we consider the case $l>0$. 
Eq.~\eqref{tE2} tells that if $M$ is negative, $\frac{E_\mathrm{rel}}{\mu}$ is positive and consequently the only possible orbit is a hyperbola. 
On the other hand, if $M$ is positive, $\frac{E_\mathrm{rel}}{\mu}$ can be negative or positive.  
For $\frac{E_\mathrm{rel}}{\mu}>0$, the orbit becomes a hyperbola, 
for $\frac{E_\mathrm{rel}}{\mu}=0$, the orbit becomes a parabola, 
for $0> \frac{E_\mathrm{rel}}{\mu} > - \frac{G^2\mu^2 M^2}{2l^2}$, the orbit becomes an ellipse 
and for $\frac{E_\mathrm{rel}}{\mu} = - \frac{G^2\mu^2 M^2}{2l^2}<0$, the orbit becomes a circle. 

\begin{figure}[h]
\begin{center}

\begin{tikzpicture}[x=0.5mm, y=0.5mm]

\draw[thin] (100,80) circle [radius=60];
\draw[thin] (100,80) circle [radius=25];

\fill (100,80) circle [radius=1];
\node at (100,85) {Center of mass};

\fill (100,55) circle [radius=1];
\node at (100,50) {$m_1>0$};

\draw (100,20) circle [radius=1];
\node at (100,15) {$m_2<0$};

\draw[->, thick] (100,55) -- (100,65);
\draw[->, thick] (100,20) -- (100,30);

\end{tikzpicture}

\end{center}

\caption{
For $m_1>-m_2$, the position vector of the centre of masses $\bm{R}$ lies on the straight line connecting both masses and outside of the mass $m_1>0$. 
The mass $m_1$ receives a repulsive force from $m_2$, which acts as a centripetal force for mass  $m_1$ with respect to the centre of masses. 
Then, for $m_1>-m_2$ and $\frac{E_\mathrm{rel}}{\mu}<0$, both masses form a bound state. 
}
\label{bndstt}
\end{figure}
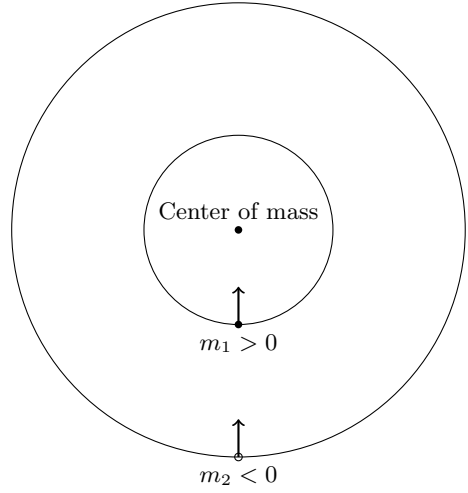

We should note that when $m_1>-m_2$, the position vector of the centre of masses $\bm{R}$ lies on the straight line connecting both masses and outside of mass $m_1>0$, as shown in FIG.~\ref{bndstt}.  
In fact, if one chooses $\bm{r}_1=(0,0,0)$ and $\bm{r}_2=(0,r,0)$ with $r>0$,  $\bm{R}=\left( 0, \frac{m_2 r}{m_1+m_2}, 0\right)$ is found. 
We should also note that $\frac{m_2 r}{m_1+m_2}$ is negative, $\frac{m_2 r}{m_1+m_2}<0$.  The mass $m_2$ exerts a repulsive force on $m_1$, and the force becomes a centripetal force for $m_1$ with respect to the centre of masses.  Hence, when $m_1>-m_2$ and $\frac{E_\mathrm{rel}}{\mu}<0$, two point masses with $m_1>0$ and $m_2<0$ form a bound state. 

We may also consider the case $l=0$. 
Eq.~\eqref{MmuRr} tells that if $M$ is positive, $\mu$ is negative, whereas if $M$ is negative, $\mu$ is positive. 
Therefore if $E_\mathrm{rel}>0$, $\frac{E_\mathrm{rel}}{\mu M}<0$, and if $E_\mathrm{rel}<0$, $\frac{E_\mathrm{rel}}{\mu M}>0$. 
This tells that in the case that $E_\mathrm{rel}>0$ and $M>0$, the solution is given by \eqref{solngtv2}, which is a bound system. 
In the case that  $E_\mathrm{rel}>0$ and $M<0$, the solution is given by \eqref{solngtv}, where the system is not bound. 
In the case that  $E_\mathrm{rel}<0$ and $M>0$, the solution is given by \eqref{solpls}, which is not a bound system. 
The case that  $E_\mathrm{rel}<0$ and $M<0$ is kinematically forbidden. 

\subsection{$m_1=-m_2=m>0$ case}

We may also consider a special case $m_1=-m_2=m>0$, where the total mass $M$ in \eqref{MmuRr} vanishes and the reduced mass $\mu$ and the position vector of the centre of mass $\bm{R}$ diverge.  Eq.~\eqref{mmtm} reduces  to
\begin{align}
\label{consv}
\dot{\bm{r}}_1 - \dot{\bm{r}}_2=\bm{v}_0 \, , \quad 
\bm{r}_1 - \bm{r}_2 = \bm{v}_0 \left(t - t_0\right)\, .
\end{align}
Here $\bm{v}_0$ is a constant vector. Then, the equations in \eqref{eom} have the following form 
\begin{align}
\label{eom3}
m \ddot{\bm{r}}_1 = - m \ddot{\bm{r}}_2 = - \frac{G m^2 \bm{v}_0}{\left| \bm{v}_0\right|^3 \left(t - t_0\right)^2}\, \, ,
\end{align}
whose solution is given by
\begin{align}
\label{slem}
\bm{r_1} = \bm{r}_2 + \bm{v}_0 \left(t - t_0\right) = \frac{G m^2 \bm{v}_0}{\left| \bm{v}_0\right|^3} \ln \frac{t - t_0}{t_0} + \bm{v}_1 t + \bm{r}_0 \, .
\end{align}
Here $\bm{v}_1$ and $\bm{r}_0$ are constants vectors of integration. 
If there are two objects such that $t<t_0$, the second equation of Eq.~\eqref{consv} implies that both objects collide with each other and they may vanish. 
The singular behaviour as $t\to t_0$ could indicate that Newton's approximations for gravity and mechanics become invalid.  
In \cite{Manfredi:2026qoz}, the above behaviours have been considered. This is the so-called runaway motion \cite{Bondi:1957zz}. 

The above calculations are, of course, not intended as a fundamental model of negative-mass particles.  As we mentioned in the introduction, in elementary mechanics, a negative inertial mass gives a kinetic energy unbounded from below, such that a point particle cannot be a stable microscopic degree of freedom. 
In this section, the sign of mass is used only as an effective weak-field parameter, which may also represent the leading gravitational response of a hypothetical extended or geometrical object that behaves, at large distances, as if it had negative mass, as discussed in \cite{Nojiri:2026ubx}. 
Within that approximation, the Newtonian two-body problem shows that positive/negative-mass systems can admit bounded relative orbits, which then provide the basis for the gravitational-wave discussion in Section~\ref{GW}.

\section{Gravitational waves}\label{GW}

A bound system of two objects generates a gravitational wave emission due to its motion. 
We now investigate and compare a system of two positive mass objects and a system consisting of one positive mass object and one NMO. 
For the system of one positive mass object and one NMO, we assume the total mass, $M=m_1 + m_2$, to be positive, but the reduced mass $\mu=\frac{m_1 m_2}{m_1 + m_2}$ is negative. 
For simplicity, we consider a circular orbit in \eqref{crcl}. 
By using \eqref{angmm2}, we find 
\begin{align}
\label{dottheta}
\dot\theta = \frac{l}{\mu r^2}=\frac{G^2 \mu^3 M^2}{l^3}\, .
\end{align}
Then the frequency $f$ of the rotation is given by 
\begin{align}
\label{frqcy}
f = \frac{\left| \dot\theta\right| }{2\pi} = \frac{G^2 \left| \mu\right|^3 M^2}{2\pi l^3} 
= \frac{\sqrt{GM}}{2\pi r^\frac{3}{2}} \, .
\end{align}
Here we used \eqref{crcl}. 
We also define $f$ to be positive. 
The relation between the frequency $f$ and the radius $r$ is nothing but Kepler's third law. 
Therefore, the expression \eqref{frqcy} is valid for both the system of two positive masses and the system of one positive mass and one NMO. 
The form of the generated gravitational wave $h$ for a binary system in a circular orbit can be obtained in linearized theory by considering just the quadrupole contribution in the multipolar expansion, which leads to \cite{Maggiore:2007ulw}
\begin{align}
\label{gw}
h \sim \frac{4G\mu}{D} \left( 2\pi r f \right)^2 \sin \left( 4\pi f t \right)
= \frac{4G^2 M\mu}{D r} \sin \left( \frac{2\sqrt{GM}}{ r^\frac{3}{2}} t \right) \ ,
\end{align}
where $D$ is the distance to the observer and we have omitted the angle dependence of the relative position of the observer with respect to the orbit's plane.  The rotation generates the oscillation of the quadrupole moment, which induces the gravitational wave. 
Eq.~\eqref{frqcy} also gives the frequency of the gravitational wave. 
The emitted energy induces a decrease in the energy $E$ of the system, and it can be expressed as 
\begin{align}
\label{emttn}
\frac{dE}{dt} = - \frac{32 G^4 \mu^2 M^3}{5r^5}\, ,
\end{align}
which is, of course, negative even for the system of one positive mass object and one NMO. 
The formula \eqref{emttn} was originally derived within the post-Newtonian approach of General Relativity for systems with positive mass and standard stress-energy. 
The gravitational wave \eqref{gw}, however, always carries positive energy and therefore $dE/dt$ is always negative. 
The emission of the gravitational wave is generated by the oscillation of the quadrupole moment, and therefore, if we know the frequency and the magnitude of the quadrupole moment, we can estimate the emitted energy. 
Therefore, the formula \eqref{emttn} can be applied even for the system including NMO(s). 

Because the radius of the circular orbit is given by \eqref{crcl}, the energy $E_\mathrm{rel}$ is expressed as 
\begin{align}
\label{Er}
E_\mathrm{rel} = - \frac{G\mu M}{2r}\, .
\end{align}
Now we identify $E$ in \eqref{emttn} with $E_\mathrm{rel}$. 
We should note that the energy $E_\mathrm{rel}$ must decrease by the emission of gravitational waves. 
For the system of two positive mass objects, due to the emission of gravitational waves, the radius $r$ becomes shorter because the energy $E_\mathrm{rel}$ must decrease. 
Then, the frequency $f$ \eqref{frqcy} becomes larger and larger. 
On the other hand, in the system with one positive mass and one NMO, because $E_\mathrm{rel}$ in (\ref{Er}) is positive, $r$ becomes larger by the emission of gravitational waves and the frequency $f$ in \eqref{frqcy} becomes smaller. By the emission of gravitational waves, the positive mass object and the NMO do not merge. 
Therefore, we can distinguish the system of two positive mass objects and the system with one positive mass and one NMO. 
By identifying $E=E_\mathrm{rel}$ and combining \eqref{emttn} and \eqref{Er},  we obtain
\begin{align}
\label{req}
\frac{G\mu M\dot r}{2r^2} = - \frac{32 G^4 \mu^2 M^3}{5r^5}\, ,
\end{align}
which can be integrated leading to
\begin{align}
\label{reqsl}
r^4 = \frac{2^8}{5} G^3 \mu M^2 \left( t _0 - t \right)\, .
\end{align}
Here, $t_0$ is a constant of integration. 
Then the frequency \eqref{frqcy} of the emitted gravitational waves has the following time dependence
\begin{align}
\label{frqcyt}
f = \frac{\sqrt{GM}}{2^4 \pi} \left\{ \frac{G^3 \mu M^2}{5} \left( t _0 - t \right) \right\}^{-\frac{3}{8}}
\, .
\end{align}
Then the gravitational wave $h$ in \eqref{gw} is given by 
\begin{align}
\label{gwt}
h \sim \frac{5^\frac{1}{4} G^\frac{5}{4} M^\frac{1}{2} \mu}{D \left\{ \mu \left( t _0 - t \right)^\frac{1}{4} \right\}} 
\sin \left( \frac{\sqrt{GM}}{2^2} \left\{ \frac{G^3 \mu M^2}{5} \left( t _0 - t \right) \right\}^{-\frac{3}{8}} t \right) \, .
\end{align}
Both the amplitude and the frequency diverge as $t\to t_0$, although the Newton approximations will be invalid there. 
In the gravitational wave for the system with two positive mass objects, the frequency and the amplitude rapidly increase when $t\to t_0$. 
On the other hand, in the behaviour of the gravitational wave for the system with one positive mass object and an NMO, the frequency and the amplitude decrease when $t$ increases. 
We should note $t>t_0$ because $\mu<0$ for the system with one positive mass object and an NMO. 
This tells us that if we find the gravitational wave where the frequency and the amplitude decrease, this might be the gravitational wave emitted from the system of one positive mass object and an NMO. 

On the other hand, the emitted energy per time in \eqref{emttn} has the following form, 
\begin{align}
\label{emttnt}
\frac{dE}{dt} = - \frac{32 G^4 \mu^2 M^3}{5\cdot 2^{10}} \left\{ \frac{G^3 \mu M^2}{5} \left( t _0 - t \right) \right\}^{-\frac{5}{4}} \, ,
\end{align}
which gives the amplitude of the gravitational wave. 

In the case of  two positive mass objects, where, of course, $\mu>0$, Eq.~\eqref{reqsl} tells that $t<t_0$, but when $t\to t_0$, the two objects collide with each other and may merge. 
When $t\sim t_0$, Newton's law of gravity must be replaced by Einstein's General Relativity, and the internal structure of the objects also becomes important. 
Eq.~\eqref{frqcyt} implies that the frequency becomes very large when $t\to t_0$ and the emitted energy expressed by Eq.~\eqref{emttnt} also diverges. 

On the other hand, in a system where one object has a positive mass and another one has a negative mass, although the total mass is positive, Eq.~\eqref{reqsl} implies that $t>t_0$. As $\mu<0$.  
Eq.~\eqref{reqsl} also tells that the distance between two objects becomes larger and larger. 
Eqs.~\eqref{frqcyt} and \eqref{emttnt} give that both the frequency and the emitted energy become smaller and smaller in time. 
The behaviours are very different from those of a system with two positive mass objects. 
Therefore, if we observe gravitational waves where the frequency and the intensity decrease, this might be the gravitational wave emitted from a system of one positive mass object and a NMO. 

\begin{figure}[h]
\begin{center}

\begin{tikzpicture}[x=1cm, y=2cm]
\begin{scope}[shift={(0,0)}]

\draw[->] (4.5,0) -- (10,0) node[right] {$t$};
\draw[->] (5,-1.5) -- (5,1.5) node[above] {$h(t)$};

\draw[thick, domain=5:9.9, samples=400]
plot (\x, { pow(10-\x,-0.25) * sin( pow(10-\x,-0.375) * \x r ) });
\end{scope}

\begin{scope}[shift={(8,0)}]

\draw[->] (4.5,0) -- (10,0) node[right] {$t$};
\draw[->] (5,-1.5) -- (5,1.5) node[above] {$h(t)$};

\draw[thick, domain=5:10, samples=400]
plot (\x, { pow(-4.8 + \x,-0.25) * sin(pow(-4.8 + \x,-0.375) * \x r ) });
\end{scope}

\end{tikzpicture}

\end{center}

\caption{
The left figure shows the behaviour of the gravitational wave for the system with two positive mass objects. 
The frequency and the amplitude rapidly increase when $t\to t_0$. 
The right figure shows the behaviour of the gravitational wave for the system with one positive mass object and an NMO. 
The frequency and the amplitude decrease when $t$ increases. 
}
\label{gws}
\end{figure}
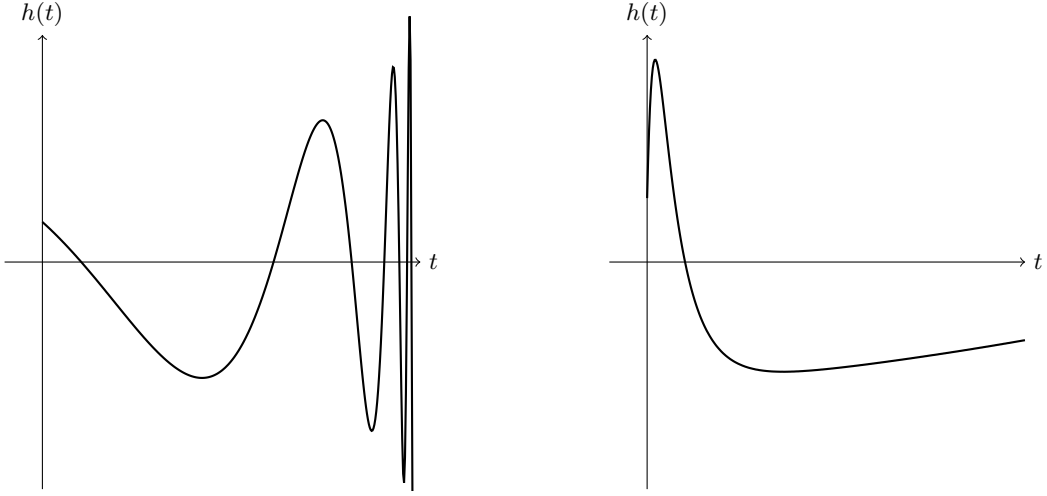

Finally, the propagating speed of gravitational waves might be identical to that of light, despite it depends on how the NMOs are generated, as in \cite{Nojiri:2026ubx}. If scalar fields source NMOs, the backreaction of the scalar fields may affect the propagation, as pointed out in \cite{Nojiri:2017hai}. 

\section{Ghost-free Ellis-Bronnikov wormhole}\label{EBwormhole}

\begin{figure}[h]
\begin{center}

\begin{tikzpicture}[x=1cm, y=1cm]

\draw[thick] (0,0) -- (0,4); 
\draw[thick] (2,1) -- (2,5); 
\draw[thick] (0,0) -- (2,1); 
\draw[thick] (0,4) -- (2,5); 

\draw[thick] (3,0) -- (3,4); 
\draw[thick] (5,1) -- (5,5); 
\draw[thick] (3,0) -- (5,1); 
\draw[thick] (3,4) -- (5,5); 

\draw[thick] (1,4) parabola bend (2.5,3) (4,4); 
\draw[thick] (1,1) parabola bend (2.5,2) (4,1); 

\draw[thick] (2.5,2.5) ellipse (0.3cm and 0.5cm); 
\draw[thick] (1.4,2.5) ellipse (0.54cm and 0.9cm); 
\draw[thick] (3.6,2.5) ellipse (0.54cm and 0.9cm); 

\draw[->] (4.3,4.3) -- (4.2,4.1);
\draw[->] (4.5,4) -- (4.3,3.8); 
\draw[->] (4.6,2.7) -- (4.4,2.6);
\draw[->] (4.5,2.2) -- (4.3,2.3); 
\draw[->] (4.3,1.5) -- (4.1,1.7); 

\draw[->] (0.8,4.1) -- (0.7,4.3); 
\draw[->] (0.7,3) -- (0.5,3.1); 
\draw[->] (0.6,2.5) -- (0.4,2.5); 
\draw[->] (0.7,2.2) -- (0.5,2.1);
\draw[->] (0.8,1.9) --  (0.6,1.7);

\end{tikzpicture}

\end{center}

\caption{
An object sucked into a wormhole that behaves as if it has positive mass is ejected from a wormhole that behaves as if it has negative mass. 
}
\label{pnwh}
\end{figure}
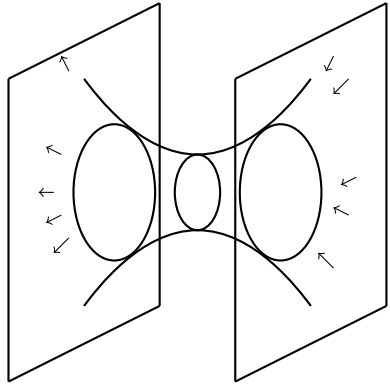

A possible candidate for a NMO is the wormhole solution found in \cite{Ellis:1973yv, Bronnikov:1973fh}, also called Ellis-Bronnikov wormhole. 
In one universe, the wormhole behaves as a positive mass object while on the other side of the throat,  the wormhole behaves as a NMO. 
The wormhole is realized by Einstein's gravity coupled with a scalar field $\phi$ with a non-canonical kinetic term, whose action is given by, 
\begin{align}
\label{EBactn}
S=\frac{1}{2\kappa^2}\int d^4 x \sqrt{-g} \left( R + 2 g^{\mu\nu} \partial_\mu \phi \partial_\nu \phi \right) \, .
\end{align}
The wormhole solution is given by \cite{Huang:2020qmn}
\begin{align}
\label{BEwh}
ds^2 =&\, - \e^{2\nu(r)}dt^2 + \e^{-2\nu(r)}dr^2 + \left(r^2 + q^2 - M^2\right)\e^{-2\nu(r)} \left( d\vartheta^2 + \sin^2\vartheta d\varphi^2 \right) \, , \nonumber \\
\phi (r) =&\,  - \frac{q\nu(r)}{M} \, , \nonumber \\
\nu(r)=&\, - \frac{2M}{\sqrt{q^2 - M^2}} \arctan \left( \frac{r}{\sqrt{q^2 - M^2}} \right) \, , 
\end{align}
Here, $q$ appears as a constant of integration and the coordinate $r$ ranges from $-\infty$ to $+\infty$, where positive $r$ may correspond to our universe and negative $r$ to another universe. 
Because $\arctan x = \frac{\pi}{2} - x^{-1} + \frac{x^{-3}}{3} + \mathcal{O}\left( x^{-5} \right)$ when $x\to +\infty$ and 
$\arctan x = - \frac{\pi}{2} + x^{-1} + \frac{x^{-3}3}{} + \mathcal{O}\left( x^{-5} \right)$ when $x\to -\infty$, 
when $\left| r \right|$ is large, in our universe $\left(r>0\right)$, $\e^{2\nu(r)}$ behaves as 
\begin{align}
\label{ournu}
\e^{2\nu(r)} \sim \e^{- \frac{2\pi}{\sqrt{q^2 - M^2}}} \left( 1 - \frac{4M}{r} + \mathcal{O}\left( r^{-2} \right) \right) \, , 
\end{align}
and in the other universe $\left(r<0\right)$, we find 
\begin{align}
\label{anthrnu}
\e^{2\nu(r)} \sim \e^{\frac{2\pi}{\sqrt{q^2 - M^2}}} \left( 1 + \frac{4M}{r} + \mathcal{O}\left( r^{-2} \right) \right) \, . 
\end{align}
The factor $\e^{\mp\frac{2\pi}{\sqrt{q^2 - M^2}}}$ could be absorbed into the redefinition of the time coordinate $t$. 
Then, Eq.~\eqref{ournu} implies that the wormhole behaves as a positive mass object in our universe, and Eq.~\eqref{anthrnu} tells that the wormhole behaves as a NMO in the other universe. 
Usual matter with positive mass may fall into the wormhole in our universe, but due to the negative mass of the wormhole in the other universe, the fallen matter is blown out by the repulsive force into another universe, as in the case of a white hole. 
Therefore, the Ellis-Bronnikov wormhole plays the role of a pump of matter at the other universe when entering the wormhole from our universe, as shown in FIG.~\ref{pnwh}. 
For electromagnetic waves or gravitational waves, if the wave is emitted in the universe where the wormhole looks to be with a negative mass, it goes through the wormhole throat and the wave emerges in the universe where the wormhole looks to have a positive mass but redshifted. 
On the contrary, the wave emitted in the universe with the wormhole with positive mass will be blueshifted if the wave appears in the universe with the wormhole with negative mass. 

Because the scalar field $\phi$ in \eqref{EBactn} is a ghost, the model is physically inconsistent. 
In order to remove the ghost, we use the formulation as in \cite{Nojiri:2024dde, Nojiri:2023dvf, Nojiri:2023zlp, Elizalde:2023rds, Nojiri:2023ztz}. 
First, we define a new scalar field $\chi$ by 
\begin{align}
\label{ch}
\phi = \phi(r=\chi)\, ,
\end{align}
that is, in the solution \eqref{BEwh}, $\chi=r$. 
For $\chi$, we add the following action with a Lagrange multiplier field $\lambda$ to the action \eqref{EBactn}, 
\begin{align}
\label{lmbdactn}
S_\lambda = \int d^4x \sqrt{-g} \lambda \left( \e^{2\nu(r=\chi)} g^{\mu\nu} \partial_\mu \chi \partial_\nu \chi - 1 \right)\, .
\end{align}
By the variation of the action $S_\chi$ with respect to $\lambda$, we obtain the following constraint, 
\begin{align}
\label{lmbdcnstrt}
0 = \e^{2\nu(r=\chi)} g^{\mu\nu} \partial_\mu \chi \partial_\nu \chi - 1 \, ,
\end{align}
which is consistent with the solution \eqref{BEwh}, $\chi=r$. 
The constraint \eqref{lmbdcnstrt} eliminates the ghost and there is always a solution $\lambda=0$ with \eqref{BEwh} \cite{Nojiri:2024dde, Nojiri:2023dvf, Nojiri:2023zlp, Elizalde:2023rds, Nojiri:2023ztz}. Note that the Ellis-Bronnikov--like wormhole can be reconstructed by following the formulation proposed in \cite{Nashed:2024jqw, Katsuragawa:2024bwm}, where an arbitrary geometry can be realized by using a non-linear $\sigma$ model whose target-space metric is identified with the Ricci curvature. Therefore, the non-linear $\sigma$ model has four scalar components in four dimensions, and the four scalar fields correspond to the spacetime coordinates. Note that in $d$ dimensions, the minimum number of scalar fields is $d$. This allows to have the wormhole \eqref{BEwh} with no need of any other extra fields, as non-linear electrodynamics. The generalization to the non-stationary case, where a time dependence is introduced in the metric, can be also easily applied to this spacetime metric, as proposed in \cite{Nojiri:2023dvf, Elizalde:2023rds, Katsuragawa:2025zcy}.

\section{Photon's orbits}\label{Phtn}

Lensing effects could provide distinctive effects on the NMOs \cite{Cramer:1994qj, Safonova:2001nd, Safonova:2001vz, Eiroa:2001zz, Safonova:2002si, Shaikh:2017zfl, Takahashi:2013jqa}. In this section, we review the lensing approach \cite{Nojiri:2026ubx} and we investigate the photon's orbit in the NMO geometry and the effects of lensing by assuming that the orbit can be given by a null orbit in a static, spherically symmetric spacetime.  Since the ADM mass is negative in the spacetime of NMOs, the anti-gravity regime could work as a gravitational concave lens. 

Let us start by the Lagrangian that describes the motion of a photon, 
\begin{align}
\label{ph1g}
\mathcal{L}= \frac{1}{2} g_{\mu\nu} \dot q^\mu \dot q^\nu = \frac{1}{2} \left( - \e^{2\nu} {\dot t}^2 + \e^{2\lambda} {\dot r}^2 + r^2 {\dot\theta}^2 + r^2 \sin^2 \theta {\dot\varphi}^2 \right) \, .
\end{align}
Here, the dot ``$\dot\ $'' expresses the derivative with respect to the affine parameter. 
Because the Lagrangian $\mathcal{L}$ does not depend on the $t$ and $\varphi$, 
there exist conserved quantities corresponding to energy $E$ and angular momentum $L$, 
\begin{align}
\label{phEg}
E \equiv&\, \frac{\partial \mathcal{L}}{\partial \dot t} = - \e^{2\nu} \dot t \, , \\
\label{phMg}
L \equiv&\, \frac{\partial V}{\partial\dot\varphi}= r^2 \sin^2 \theta \dot\varphi \, . 
\end{align}
The total energy $\mathcal{E}$ of the system should be also conserved, 
\begin{align}
\label{totalEg}
\mathcal{E} \equiv \mathcal{L} - \dot t \frac{\partial \mathcal{L}}{\partial \dot t} - \dot r \frac{\partial \mathcal{L}}{\partial \dot r} 
 - \dot\theta \frac{\partial \mathcal{L}}{\partial \dot\theta} - \dot\varphi \frac{\partial \mathcal{L}}{\partial \dot\varphi} = \mathcal{L} \, . 
\end{align}
In the case of a photon, whose geodesic is null, we require $\mathcal{E}=\mathcal{L}=0$. 

Without loss of generality, we consider the orbit on the equatorial plane with $\theta=\frac{\pi}{2}$. 
Then, the condition $\mathcal{E}=\mathcal{L}=0$ gives, 
\begin{align}
\label{geo1g}
0= - \frac{E^2}{2} + \frac{1}{2} {\dot r}^2 + \frac{L^2 a(r)}{2r^2} \, .
\end{align}
For simplicity, we consider the case of a Schwarzschild-type metric
\begin{align}
\label{Schw}
\e^{2\nu(r)}=\e^{-2\lambda(r)}=1 - \frac{2M}{r}\, .
\end{align}
If $M$ is negative, the spacetime is given by an NMO. 

This system is analogous to the classical dynamical system with a potential $U(r)$ given by
\begin{align}
\label{geo2g}
0 =\frac{1}{2} {\dot r}^2 + U(r)\, , \quad U(r) \equiv \frac{L^2 \e^{2\nu(r)}}{2r^2} - \frac{E^2}{2} \, .
\end{align}
If there is a circular orbit, where $\dot r=0$, the equations $U(r)= U'(r)=0$ must have a solution. 
We find, however, 
\begin{align}
\label{phg1}
U(r) =&\, \frac{L^2}{2r^2} \left( 1 - \frac{2M}{r} \right) - \frac{E^2}{2} \, , \\
\label{phg2}
U'(r) =&\, - \frac{L^2}{r} \left( \frac{1}{r^2} - \frac{3M}{r^3} \right) \, , 
\end{align}
Eq.~\eqref{phg2} has no solution if $M$ is negative, and therefore there is no circular orbit of a photon, nor a photon sphere when $M<0$. 

For Eq.~\eqref{geo2g} with \eqref{phg1}, the radius of the turning point, where $\dot r =0$ is given by 
\begin{align}
\label{trnpnt}
0 = U(r) = \frac{L^2}{2r^2} \left( 1 - \frac{2M}{r} \right) - \frac{E^2}{2} \, .
\end{align}
which is a cubic algebraic equation, which can be solved. 
\begin{align}
\label{slcbc}
r= \left( - q + \sqrt{\Delta} \right)^\frac{1}{3} 
+ \left( - q - \sqrt{\Delta} \right)^\frac{1}{3} \, , \quad 
q\equiv  \frac{L^2M}{E^2}\, , \quad 
\Delta 
= \frac{L^4 M^2}{E^4} - \frac{L^6}{27 E^6}
\, .
\end{align}
Here we choose a real branch for $\left(\ \right)^\frac{1}{3}$ although there are two complex branches besides the real one. 
Becasue $q^2 > \Delta$ and $-q>0$, as long as $\Delta\geq 0$, Eq.~\eqref{slcbc} is a unique and positive solution of $r$, which holds even if $\Delta<0$ because the second term in Eq.~\eqref{slcbc} is the complex conjugate of the first term. We should note that because Eq.~\eqref{phg2} tells that $U'(r)<0$, that is, $U(r)$ is a monotonically decreasing function of $r$ as long as $r>0$, 
Eq.~\eqref{trnpnt} has a unique real and positive solution of $r$, which is the minimum radius or the turning point of the orbit.

As in classical mechanics, the solution of \eqref{geo2g} is given by 
\begin{align}
\label{orbit}
t = \int \frac{dr}{\sqrt{-2U(r)}} = \int \frac{dr}{\sqrt{E^2 - \frac{L^2}{r^2} \left( 1 - \frac{2M}{r} \right)}} \, ,
\end{align}
or because Eq.~\eqref{phMg} with $\theta=\frac{\pi}{2}$ tells $\frac{d\varphi}{dt} = \frac{L}{r^2}$, we rewrite \eqref{geo2g} with \eqref{phg1} in the following form, 
\begin{align}
\label{geo2gB}
0 =\frac{L^2}{2r^4} \left(\frac{dr}{d\varphi} \right)^2 + \frac{L^2}{2r^2} \left( 1 - \frac{2M}{r} \right) - \frac{E^2}{2} \, ,
\end{align}
which can be integrated to give
\begin{align}
\label{orbitph}
\varphi = \int \frac{dr}{r^2 \sqrt{\frac{E^2}{L^2} - \frac{1}{r^2} \left( 1 - \frac{2M}{r} \right)}} \, , 
\end{align}
which describes the shape of the orbit. 
An asymptotically hyperbolic curve could give the orbit for NMOs, where $M$ is negative.  Therefore, the NMO can be regarded as a gravitational concave lens.
If the NMO exists in front of the light source(s), for example, a small NMO in front of the sun, the object could be observed as a dark spot. 
If there are many light sources behind the NMO, we can observe the deformed distribution as observed in \cite{Cramer:1994qj, Safonova:2001nd, Safonova:2001vz, Eiroa:2001zz, Safonova:2002si, Shaikh:2017zfl, Takahashi:2013jqa}.

\section{Ray-tracing and shadows}\label{Shadows}

\begin{figure}[t!]
\centering
\includegraphics[scale=0.6]{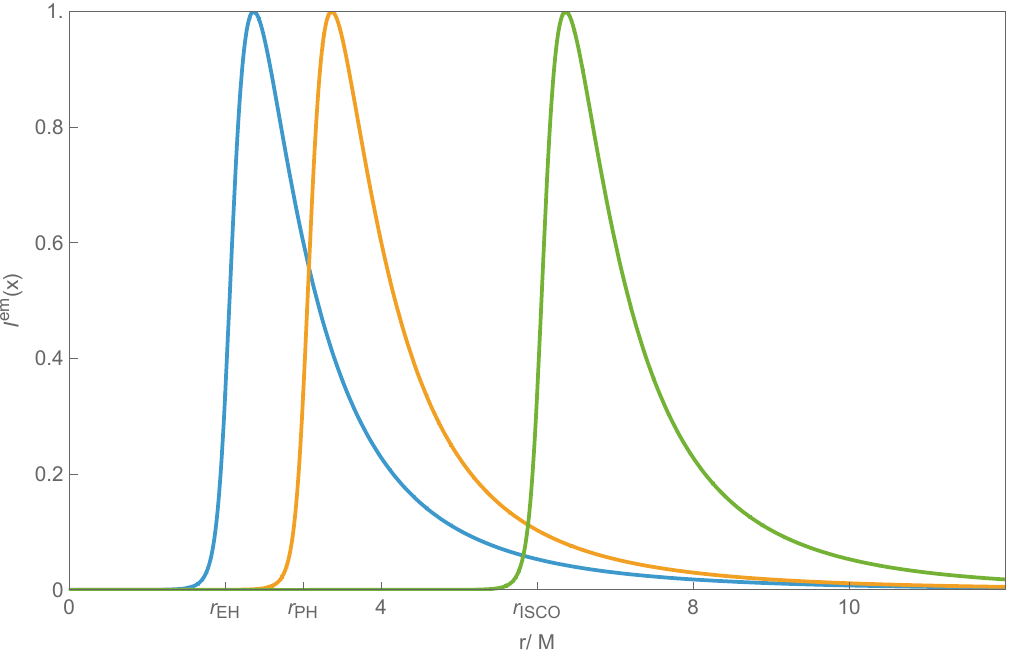}
\caption{The intensity profile \eqref{IntensityProfile} with $\gamma=-2$, $\sigma=1/4$ and $\tilde\mu=2$ (blue),  $\tilde\mu=3$ (orange) and $\tilde\mu=6$ (green), which refer to the location of the event horizon, the photon sphere and the ISCO for a Schwarzschild black hole.}
\label{Fig_Profiles}
\end{figure}

Let us now analyse the optical appearance of such a type of wormhole. 
To do so, we use the so-called ray-tracing technique that integrates the photon trajectory from the observer screen towards the object. 
The motion is just described by the geodesic equation \eqref{geo1g}, which can be rewritten in the form:
\begin{align}
\dot{r}^2=\frac{1}{b^2}-V(r)\, ,
\label{eqradialPhotons}
\end{align}
where the parameter $b$ is the so-called impact parameter $b=\frac{L}{E}=\frac{g_{\theta\theta}\dot{\phi}}{g_{tt}\dot{t}}$ and it describes each photon trajectory, whereas the effective potential is given by:
\begin{align}
V(r)=\frac{g_{tt}}{g_{\theta\theta}}\, .
\label{Photons_Potential}
\end{align}
\begin{figure*}[htb]
\centering
\includegraphics[width=0.8\linewidth]{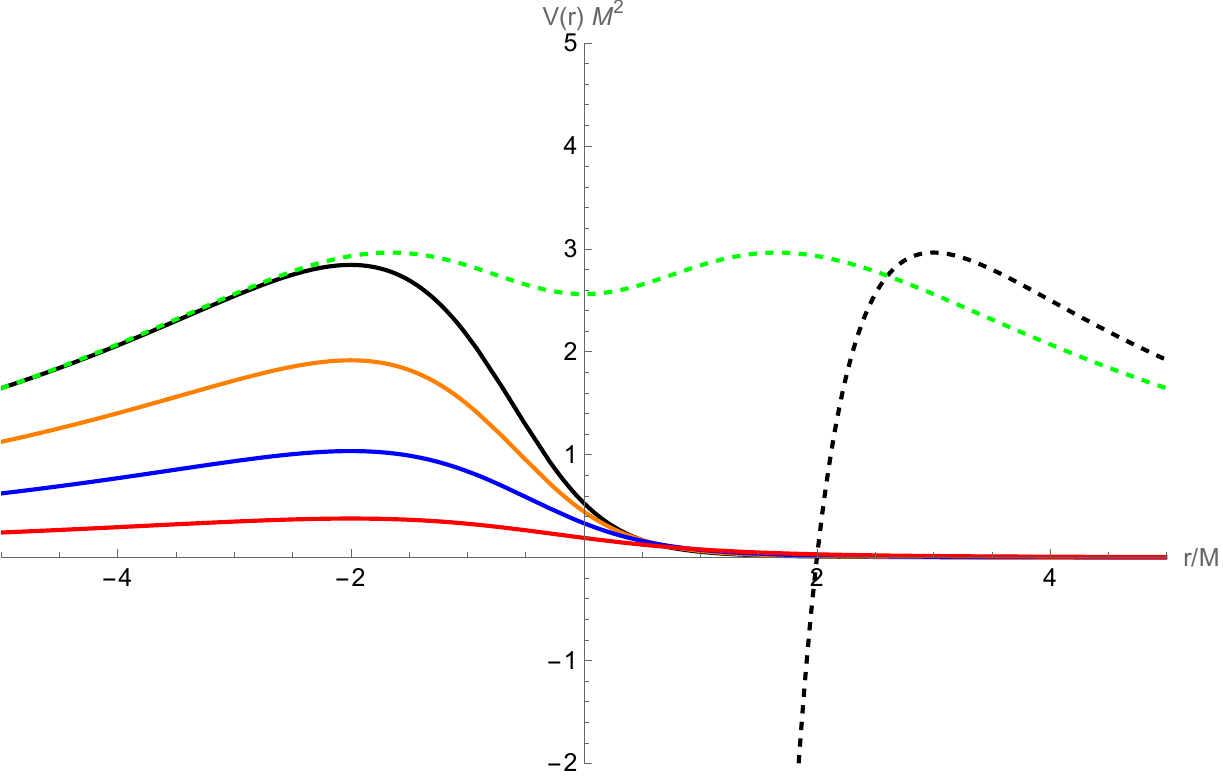}
\caption{The effective potential for photons \eqref{Photons_Potential}. 
The curves describe the Schwarzschild case (dashed black line), the Simpson-Visser wormhole with $a/M=5/2$ (dashed green line) and the wormhole \eqref{BEwh} for $q/M=1.7$ (black), $q/M=1.8$ (orange), $q/M=2$ (blue), $q/M=2.5$ (red). Note that to visualise all the potentials better, the scales of the Schwarzschild black hole and for the Simpson-Visser wormhole are increased by a factor  $\times80$.}
\label{Fig_Photons_Potential}
\end{figure*}
Here, $g_{tt}$ and $g_{\theta\theta}$ stand for the corresponding components of the metric, which are assumed to be static and spherically symmetric. 
The shape of the potential is shown in Fig.~\ref{Fig_Photons_Potential}, where one can note the asymmetry between both sides of the throat for the wormhole \eqref{BEwh}. 
Also, the effective potential for the Schwarzschild black hole is depicted, which shows a pole at the origin as expected due to the presence of a singularity. We have also included the Simpson-Visser wormhole, whose metric can be expressed as follows \cite{Simpson:2018tsi},
\begin{align}
ds^2=-\left(1-\frac{2M}{\sqrt{r^2+a^2}}\right)dt^2+\left(1-\frac{2M}{\sqrt{r^2+a^2}}\right)^{-1}dr^2+\left(r^2+a^2\right) d\Omega^2\ .
\label{Simpson-Visser_metric}
\end{align}
One should also note that the wormhole potential exhibits an unstable circular orbit for photons just in one of the sides of the throat, while it does not at the other side. 
The equation of the trajectory in the equatorial plane \eqref{orbitph} can be expressed as:
\begin{align}
\left(\frac{d\phi}{dr}\right)=\pm \frac{b}{g_{\theta\theta}\sqrt{1-\frac{b^2g_{tt}}{g_{\theta\theta}}}}\, ,
\end{align}
\begin{figure}[t!]
\centering
\includegraphics[scale=0.66]{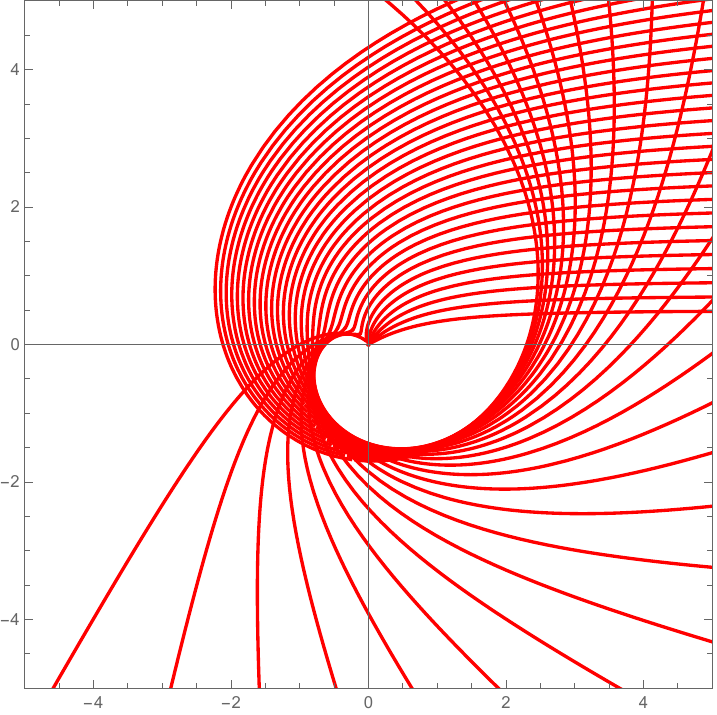}
\caption{Light trajectories in the equatorial plane for the wormhole \eqref{BEwh} with $q/M=1.8$ and an interval of the impact parameter $b\in (0,10)$.}
\label{Fig_Ray-tracing}
\end{figure}
In Fig.~\ref{Fig_Ray-tracing}, a set of trajectories is simulated for a particular range of the impact parameter. 
Then, the aim is to use where photons cross a hypothetical accretion disk around the wormhole in order to simulate the observed luminosity. 
By assuming the accretion disk to be located perpendicular to the line of sight of the observer, which in Fig.~\ref{Fig_Ray-tracing} would refer to the vertical axes, we can compute the observed luminosity. 
To do so, one should note that for some values of the impact parameter, photon trajectories cross the vertical axes, and consequently the accretion disk, more than once, which implies an additional brightness in the observer screen due to the extra photons approaching with the same impact parameter but emitted at different locations of the accretion disk. 
To compute the total luminosity, one has to sum over all the contributions:
\begin{align}
\left. I_\mathrm{obs}=\sum_{m} g_{tt}(r)^2 I^\mathrm{em}(r)\right|_{r=r_m(b)}\, ,
\label{intensity}
\end{align}
\begin{figure}[t!]
\centering
\includegraphics[scale=0.53]{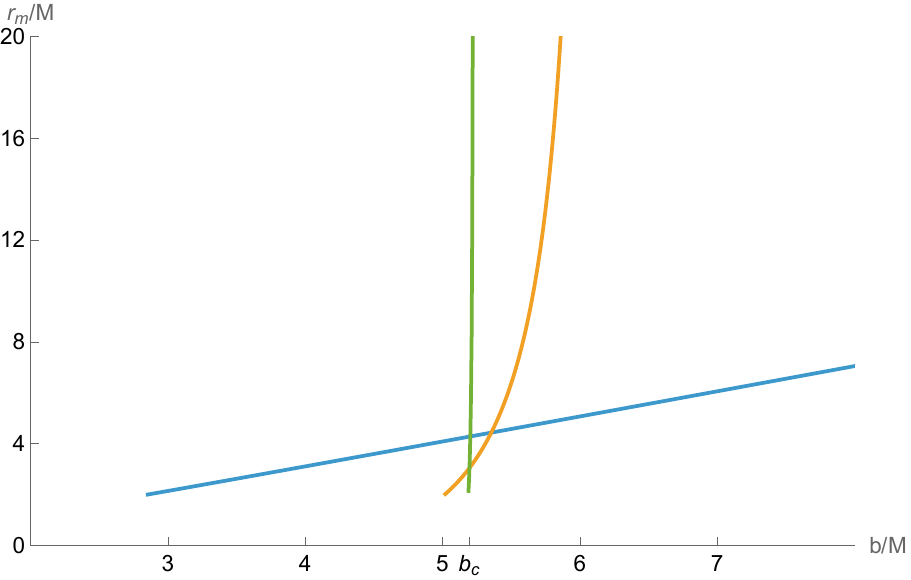}
\includegraphics[scale=0.53]{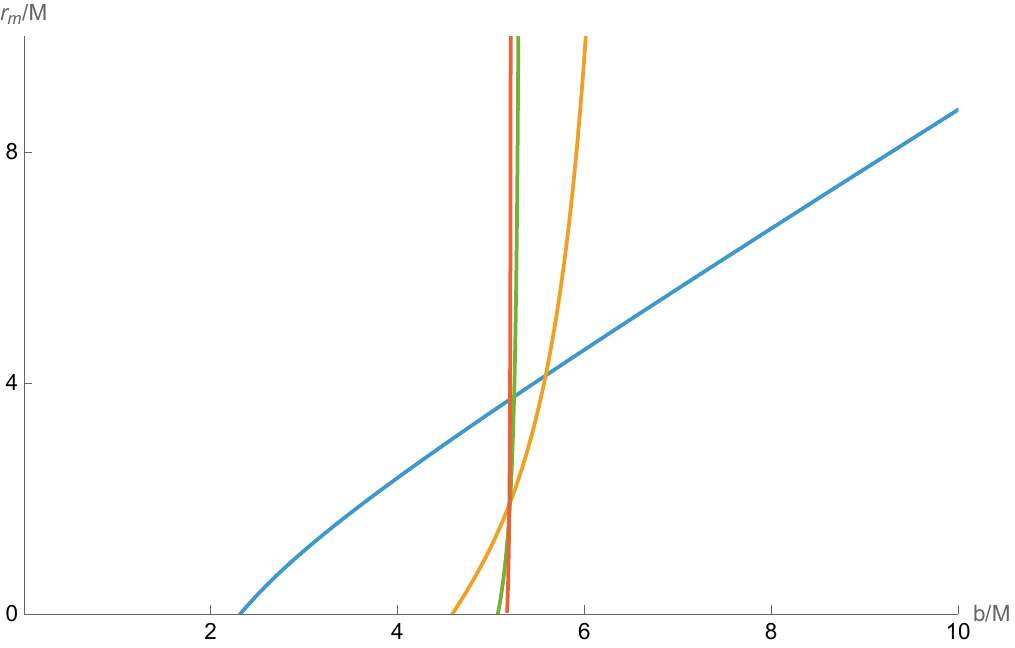}
\includegraphics[scale=0.53]{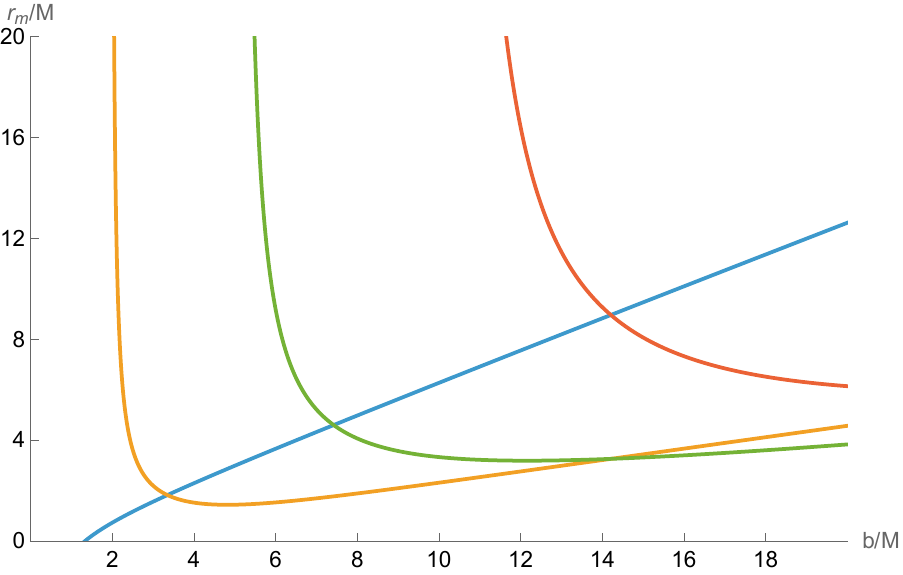}
\caption{Transfer functions for the Schwarzschild black hole (left panel), the Simpson-Visser wormhole with $a/M=5/2$ (central panel) and the wormhole \eqref{BEwh} with $q/M=1.8$ (right panel). 
The plots show the curves for the direct emission (blue), the lensed emission (orange) and higher order cases (green and red).}
\label{Fig_Transfer}
\end{figure}
where $I(r)$ is the luminosity profile of the accretion disk and $r_m(b)$ are the so-called transfer functions that characterise the location of photon emission as a function of the impact parameter. 
Such transfer functions are depicted in Fig.~\ref{Fig_Transfer}, where the subscript $m$ refers to the direct emission ($m=0$), the lensed emission ($m=1$) and higher order emissions ($m\geq 2$), which just collect the locations where a photon with a particular impact parameter might intersect the central object according to its trajectory. 
Moreover,  for describing the emission of the accretion disk, we are assuming a particular intensity profile that, despite its simplicity, might describe the emission of the accretion disk effectively according to previous gravitomagnetic-hydrodynamics simulations \cite{Paugnat:2022qzy}. 
Such a profile can be cast as follows:
\begin{align}
I^\mathrm{em}(r,\tilde\mu,\sigma,\gamma)=\frac{\e^{-\frac{1}{2}\left[\gamma+\mathrm{arcsinh}\left(\frac{r-\tilde\mu}{\sigma}\right)\right]^2}}{\sqrt{(r-\tilde\mu)^2+\sigma^2}}\, .
\label{IntensityProfile}
\end{align}
Here, $\tilde\mu$, $\sigma$ and $\gamma$ are free parameters that describe the position of the emission peak, its width and the size of the central region, respectively. 
Fig.~\ref{Fig_Profiles} gives a representation of such a type of profile. 
For the sake of this work, we are considering for every simulation the same width and size for the intensity but varying the location of the peak, as shown in Fig.~\ref{Fig_Profiles}. 
Note also that for the case of the NMO wormhole \eqref{BEwh}, the intensity profile is assumed to have its peak at the throat $r=0$, at the location of the ISCO $r=0.49$, but also at $r=3M$ and at $r=6M$, where the photon sphere and the ISCO are located for the Schwarzschild black hole. This is because the wormhole \eqref{BEwh} has no circular orbit for photons for $r>0$, as it is inferred in the shape of the potential in Fig.~\ref{Fig_Photons_Potential}, while ISCO is very close to the central object $r/M=0.49$, so that in order to guarantee a better comparison, we also include such simulations.\\ 

\begin{figure*}[htb]
    \centering
    \includegraphics[width=0.58\linewidth]{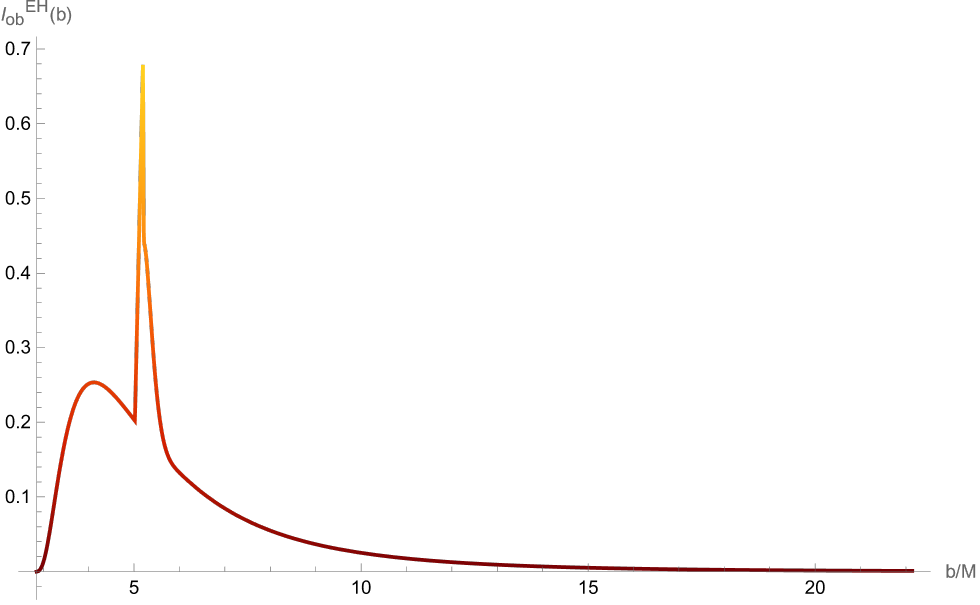}
        \includegraphics[width=0.4\linewidth]{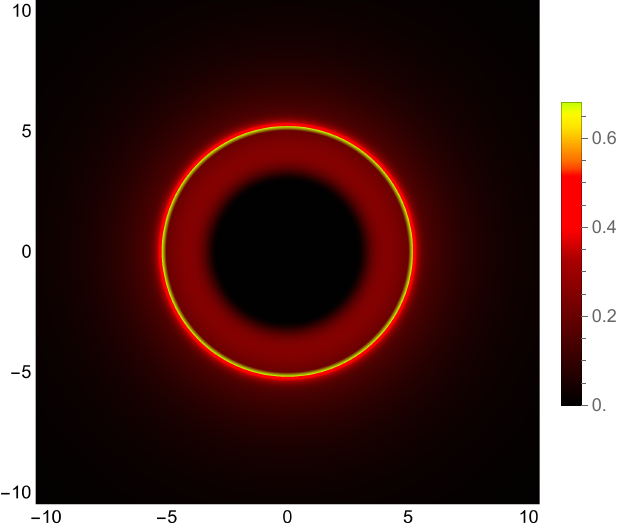}
\includegraphics[width=0.58\linewidth]{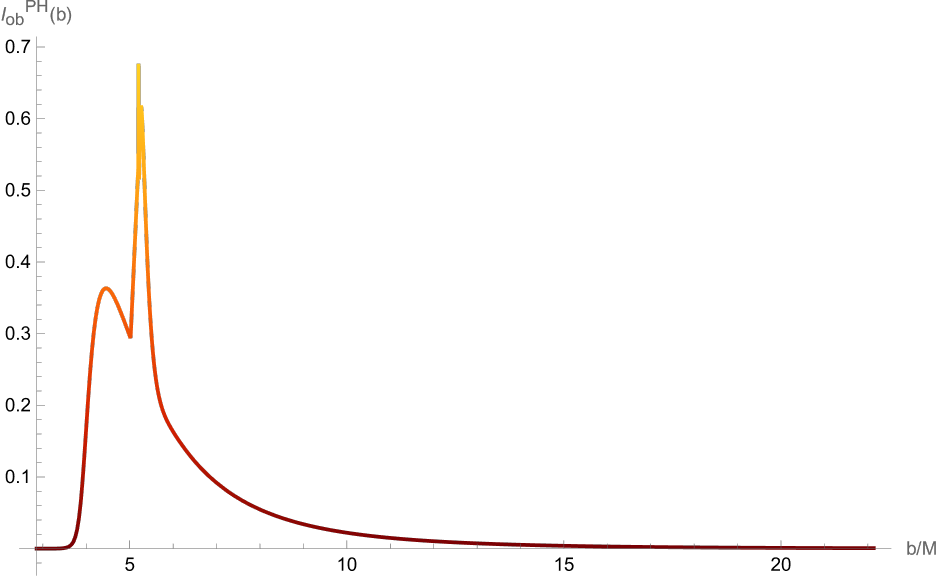}
        \includegraphics[width=0.4\linewidth]{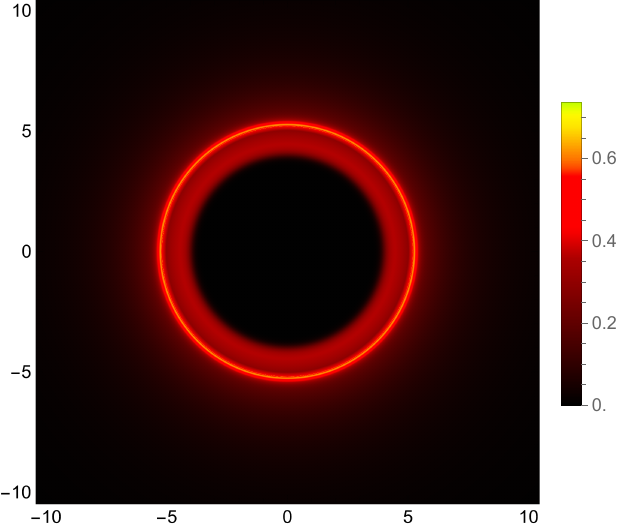}
        \includegraphics[width=0.58\linewidth]{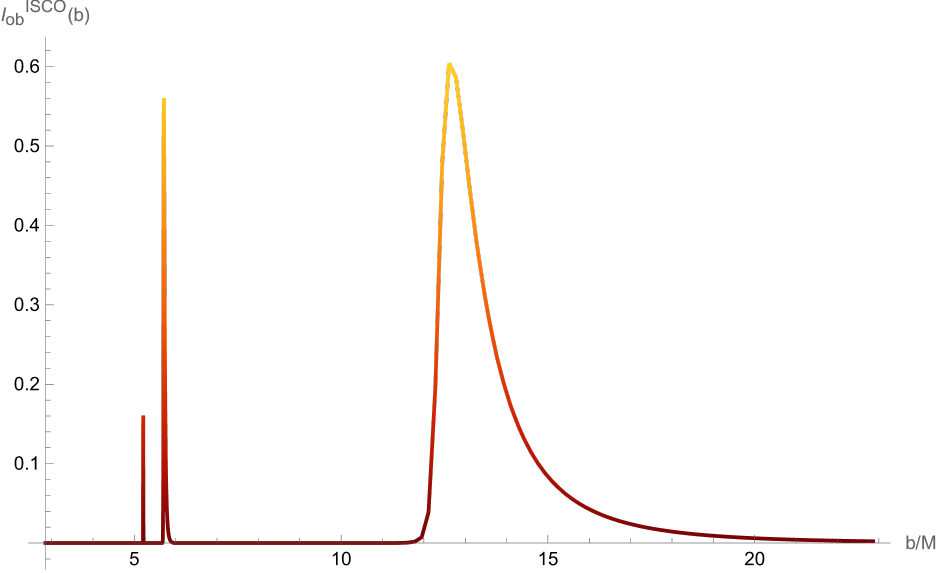}
           \includegraphics[width=0.4\linewidth]{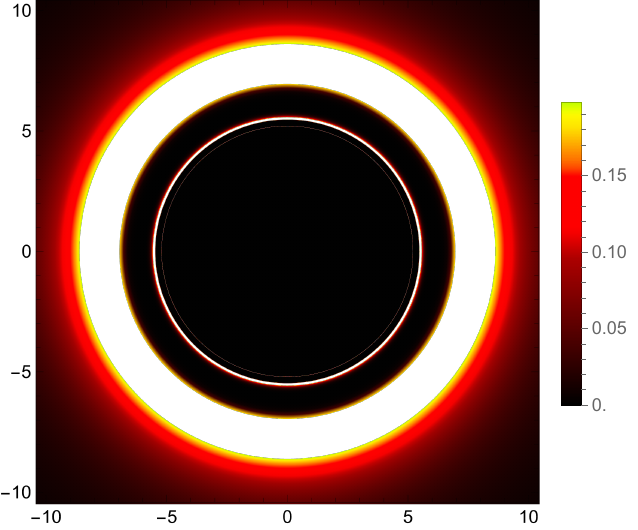}
    \caption{Observed intensities (left panels) and the shadows (right panels) for a Schwarzschild black hole. 
    From top to the bottom: the peak of the luminosity located at $r=2M$ (Event horizon), at $r=3 M$ (photon sphere) and at $r=6 M$ (ISCO).}
    \label{Fig_Optical_Appe_SCHW}
\end{figure*}

\begin{figure*}[htb]
    \centering
    \includegraphics[width=0.58\linewidth]{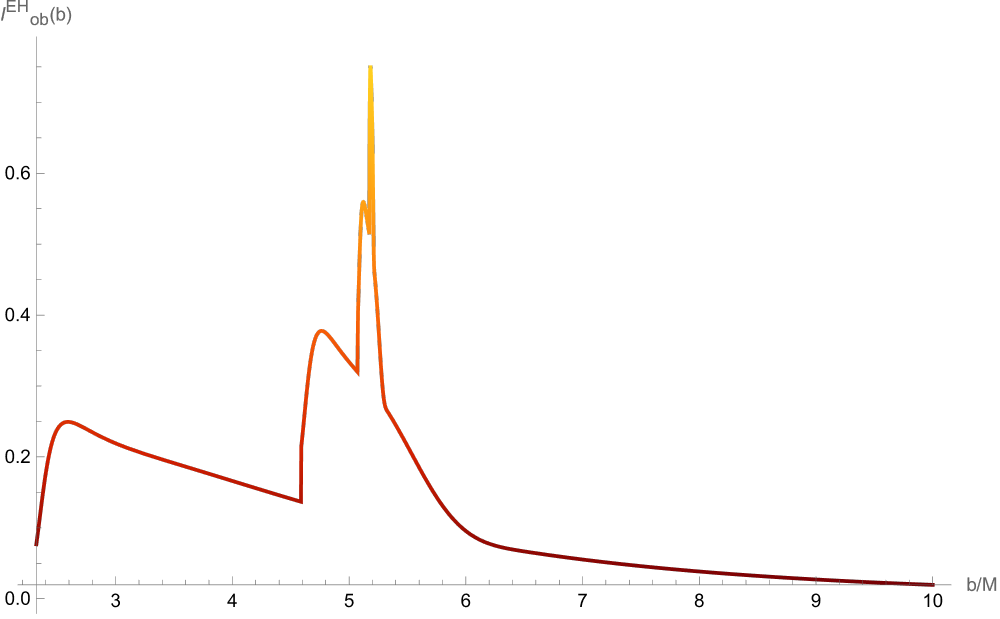}
        \includegraphics[width=0.4\linewidth]{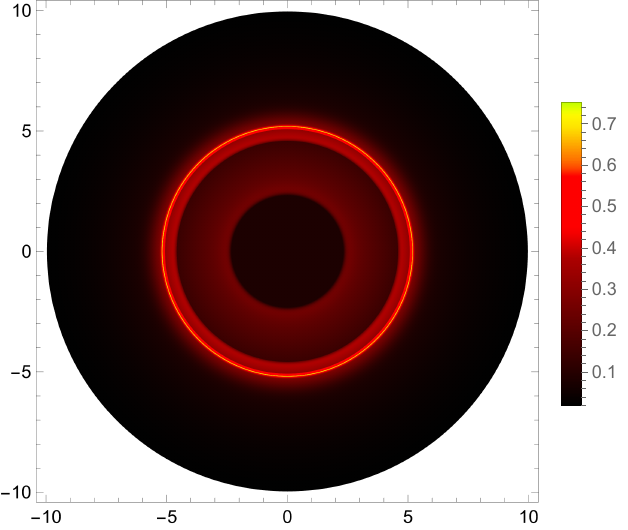}
\includegraphics[width=0.58\linewidth]{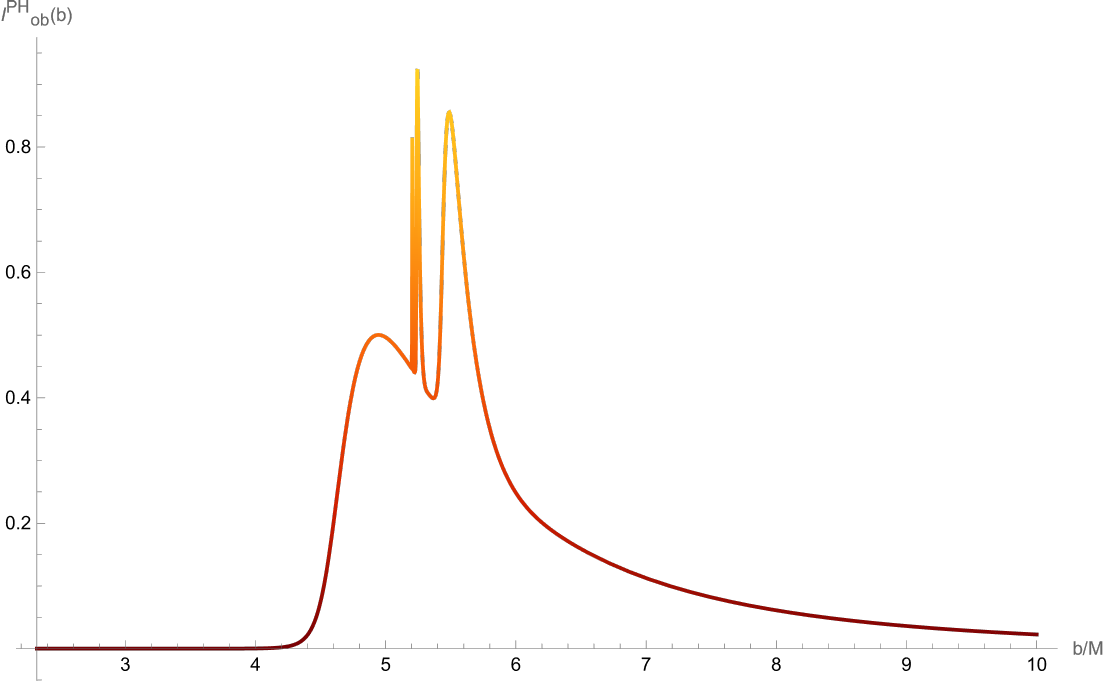}
        \includegraphics[width=0.4\linewidth]{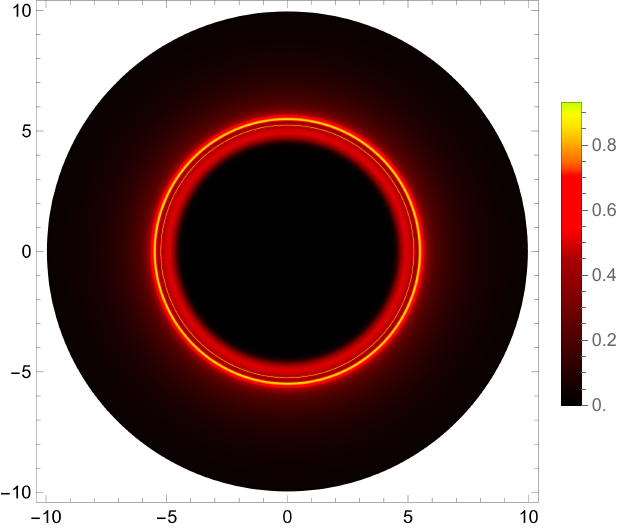}
        \includegraphics[width=0.58\linewidth]{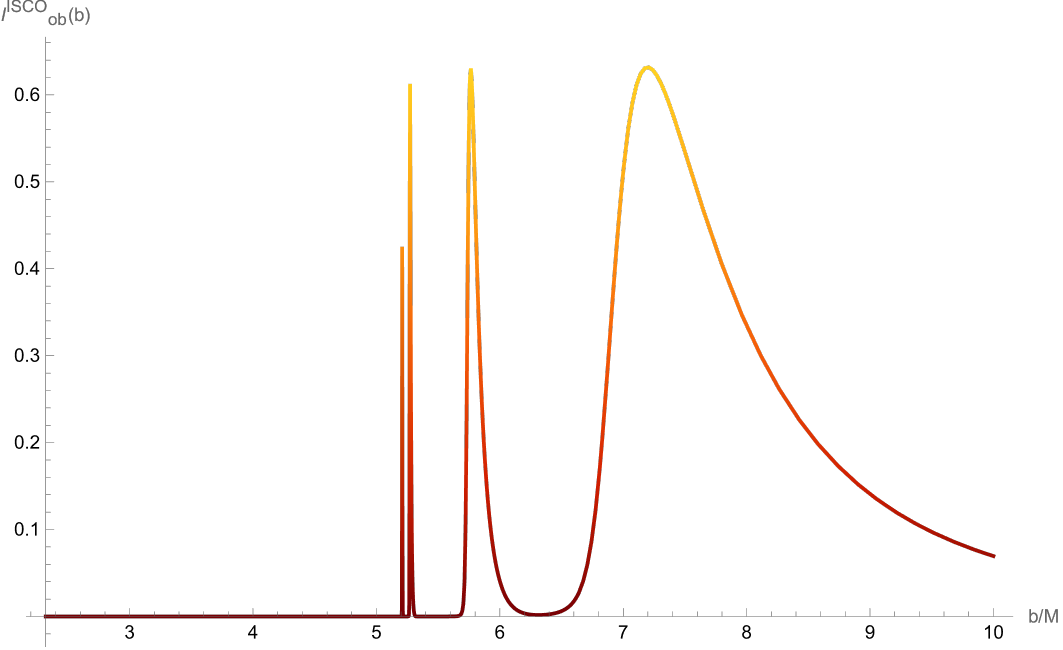}
           \includegraphics[width=0.4\linewidth]{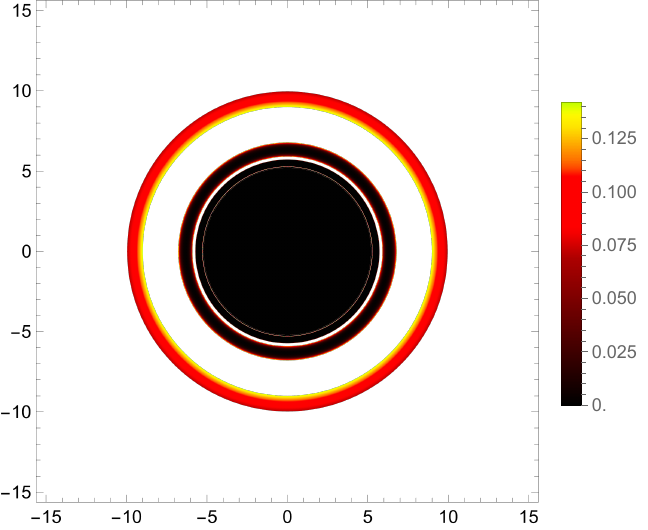}
    \caption{Observed intensities (left panels) and the shadows (right panels) for a Simpson-Visser wormhole with $a=5/2$. 
    From top to the bottom: the peak of the luminosity located at $r=0$ (the throat), at $r= 3M$ (photon sphere) and at $r=5.45 M$ (ISCO).}
    \label{Fig_Optical_Appe_Simpson}
\end{figure*}

\begin{figure*}[htb]
    \centering
    \includegraphics[width=0.58\linewidth]{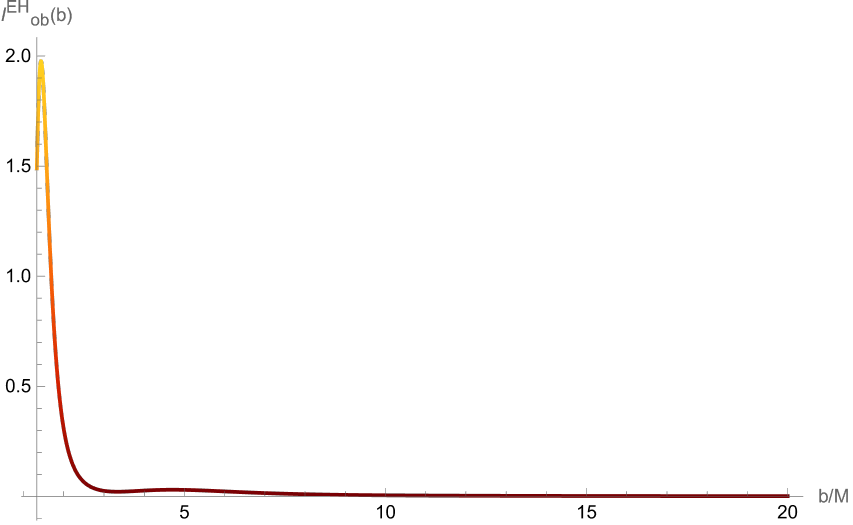}
        \includegraphics[width=0.4\linewidth]{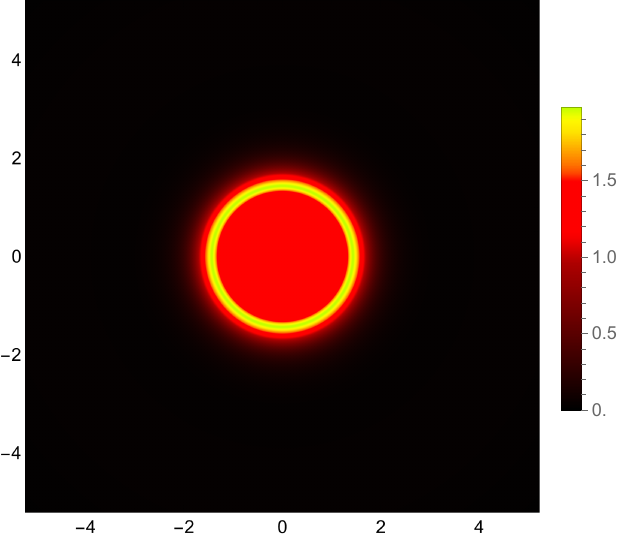}
 \includegraphics[width=0.58\linewidth]{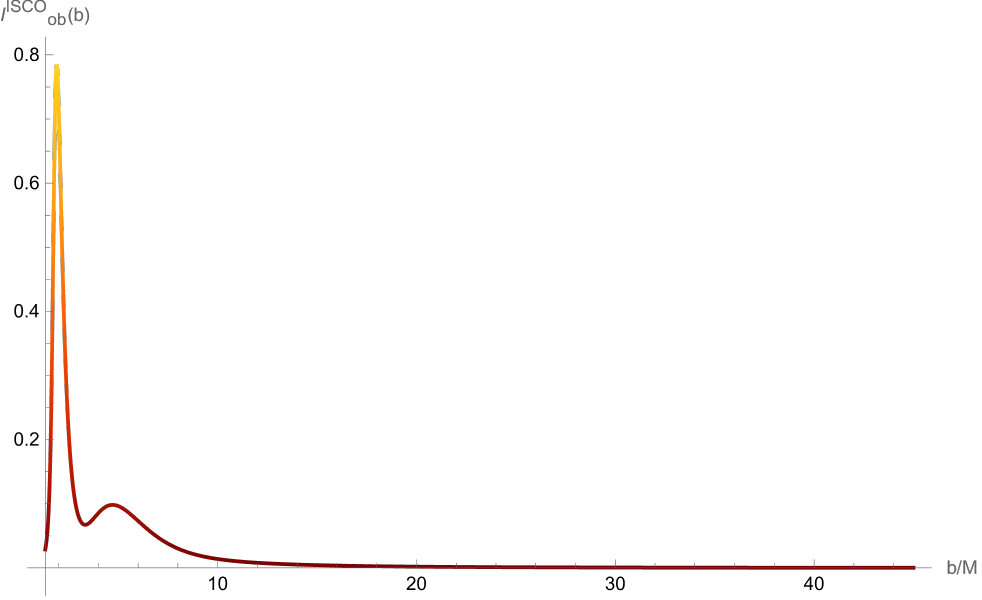}
           \includegraphics[width=0.4\linewidth]{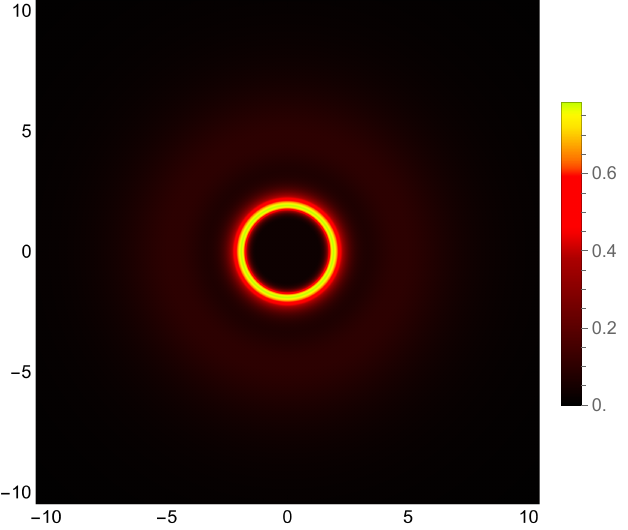}
          \caption{Observed intensities (left panels) and the shadows (right panels) for the wormhole black hole \eqref{BEwh} with $q/M=1.8$. 
From top to the bottom, the locations of the peaks are at $r=0$ (the throat) and $r=0.49M$ (ISCO).}
    \label{Fig_Optical_Appe_WH}
\end{figure*}

\begin{figure*}[htb]
    \centering
   \includegraphics[width=0.58\linewidth]{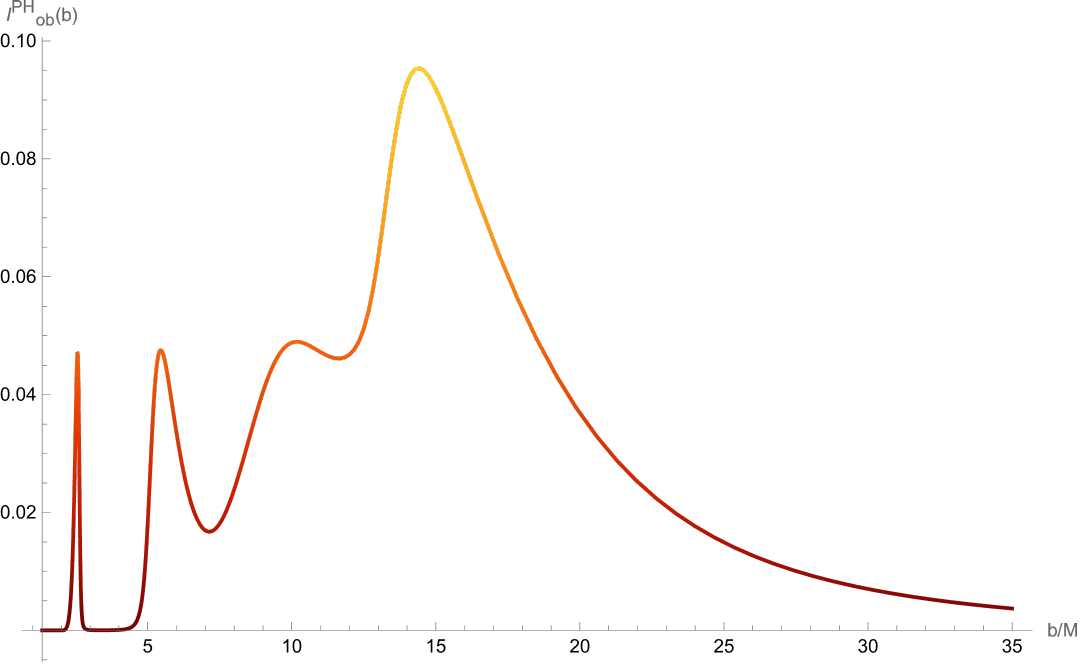}
        \includegraphics[width=0.4\linewidth]{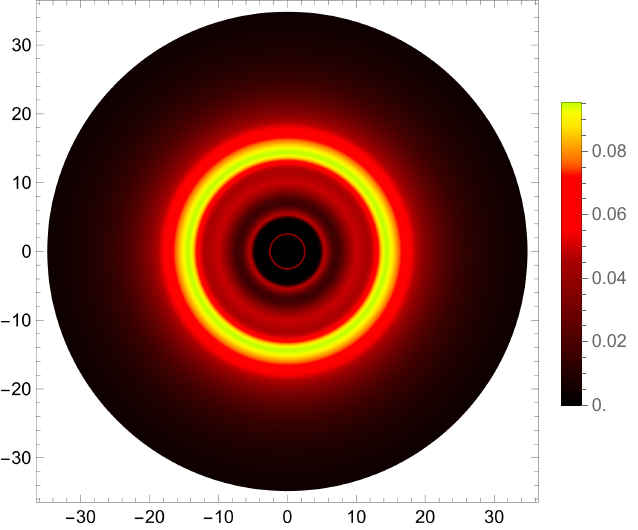}
         \includegraphics[width=0.58\linewidth]{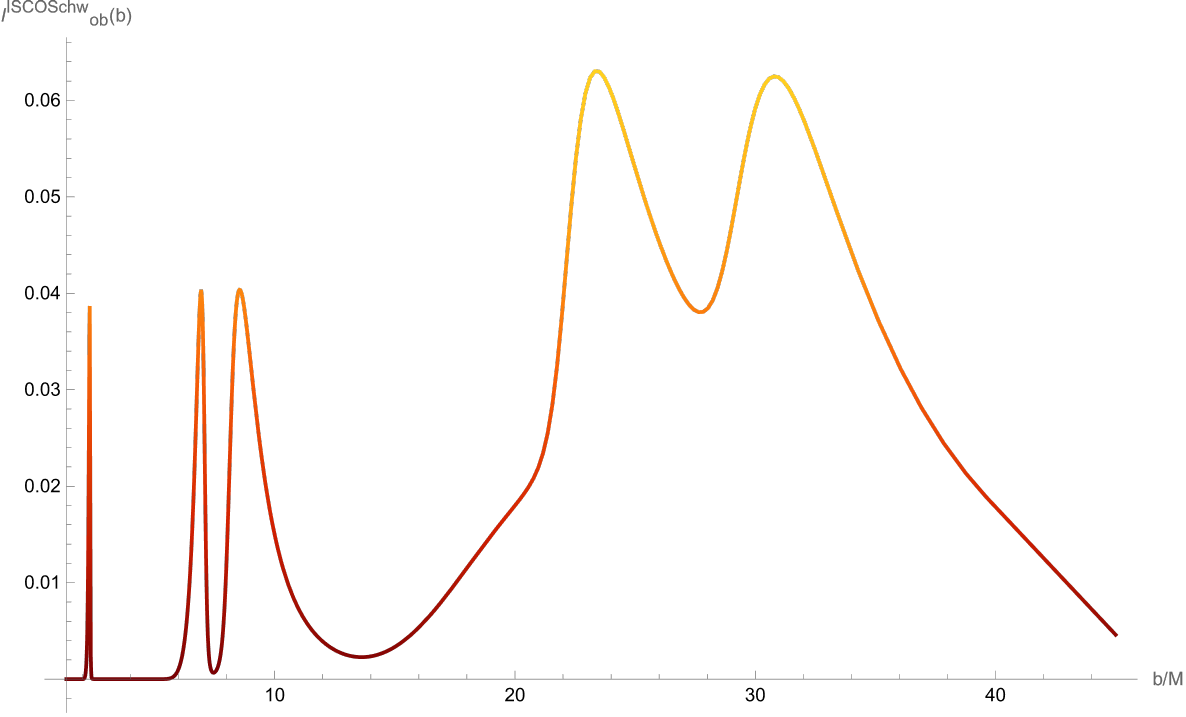}
           \includegraphics[width=0.4\linewidth]{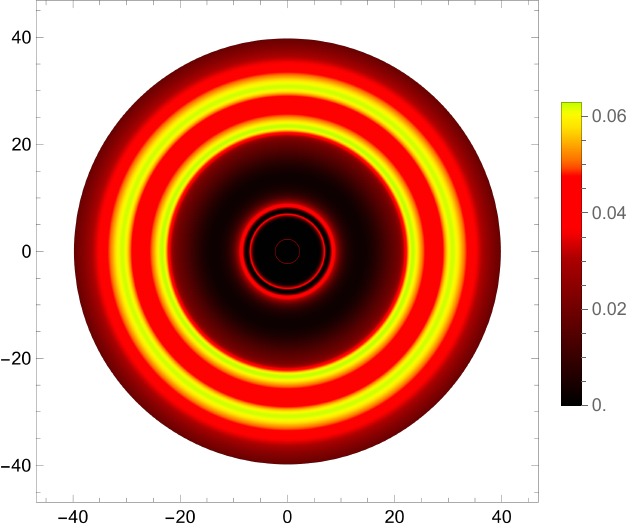}
          \caption{Observed intensities (left panels) and the shadows (right panels) for the wormhole black hole \eqref{BEwh} with $q/M=1.8$. 
From top to the bottom, the locations of the peaks are at $r=3 M$ (Schwarzschild photon sphere) and $r=6M$ (Schwarzschild ISCO).}
    \label{Fig_Optical_Appe_WH1}
\end{figure*}

Then, in Figs.~\ref{Fig_Optical_Appe_SCHW}, \ref{Fig_Optical_Appe_Simpson}, \ref{Fig_Optical_Appe_WH} and \ref{Fig_Optical_Appe_WH1}, the luminosity as collected by a far away observer is depicted for the Schwarzschild black hole, the Simpson-Visser wormhole~\cite{Simpson:2018tsi} and the wormhole~\eqref{BEwh} with $q/M=1.8$. 
As such figures point out, the photon rings' substructure is much more complex in the case of the NMO wormhole when the peak of the luminosity is a bit far away from the throat, not only in comparison to the Schwarzschild black hole but also when comparing to another wormhole structure. 
Such a result is a direct consequence of the asymmetry of the effective potential \ref{Fig_Photons_Potential} in the NMO wormhole, which exhibits a maximum (a circular orbit for photons) on the other side of the throat, which affects what an observer would collect on the screen at the positive side of the throat. Nevertheless, when the peak of the luminosity is close to the throat, and particularly to the ISCO, differences are not so notorious despite the observed luminosity reaching the throat in comparison to the Simpson-Visser wormhole, where it does not. 

\clearpage

\section{Summary and Discussion}\label{SD}

In this paper, we have investigated systems containing NMOs. By starting with a tutorial review of the classical mechanics of a system including NMOs in Section~\ref{udrgrdtmch}, we have found that a system consisting of one object with positive mass and one NMO, a bound state exists even though the force exerted by the NMO on the object with positive mass is repulsive. 
Unlike a standard system consisting of two objects with positive mass, the emission of gravitational waves from this system exhibits a decrease in frequency and amplitude over time, as shown in Section~\ref{GW}. In Section~\ref{EBwormhole}, we propose a formulation to exclude the ghosts that arise in the model realizing the Ellis-Boronnikov wormhole~\cite{Ellis:1973yv, Bronnikov:1973fh}, a candidate of an NMO. 

After reviewing the lensing effect by the NMOs in Section~\ref{Phtn},  by using the so-called ray-tracing technique, we have shown the numerical results for the observed image of NMOs or the Ellis-Bronnikov wormhole, numerically in Section~\ref{Shadows}. 
By computing the luminosity and optical appearance of the NMO wormhole in comparison to the canonical Schwarzschild black hole and the Simpson-Visser wormhole, large differences arise in the intensity and the images as a result of the asymmetric potential for the photons' trajectories that characterises such NMO objects, showing a much more complex photon rings substructure in some cases. 
One should note that the asymmetry potential for photon trajectories depicted in Fig.~\ref{Fig_Photons_Potential} shows that on one side of the throat ($r<0$) it has a similar shape as other more usual objects, such as the Simpson-Visser spacetime and even, up to some limit, to the Schwarzschild black hole. 
This means that for $r<0$, photons will follow similar paths as in Simpson-Visser spacetime. 
Indeed, differences arise when dealing with the other side of the throat ($r>0$), where the absence of a maximum in the photon's potential avoids possible locations for any circular photon orbits. 
In addition, the potential remains constant and small till one approaches the central object closely, which produces a high deflection on the photons that are emitted by the accretion disk far from the central object, while its luminosity decreases very fast when the peak is close to the throat. 
The result in the optical appearance of the object is an interesting structure of photon rings surrounding the object that shows large differences with other spacetimes. 
Firstly, when the peak of the luminosity is located at the wormhole throat, the observed intensity also reaches the throat, contrary to the Simpson-Visser wormhole \cite{Guerrero:2021ues}. 
Moreover, when the peak is located further from the throat, one finds that those photons experience high deflections, leading to an observed luminosity that contains several peaks that then give a complex photon ring structure in comparison to the other two spacetimes. 

Although one might not expect the existence of such objects, this analysis intends to give a smoking gun that might be used in the future to infer the existence of objects beyond the Kerr paradigm and the current fundamental physics.

\section*{Acknowledgements}

DS-CG is supported by the Spanish project PID2024-157196NB-I00 funded by MICIU/AEI/10.13039/501100011033  (“ERDF A way of making Europe”, “PGC Generaci\'on de Conocimiento”) ; and the financial support by the Department of Education, Junta de Castilla y Le\'{o}n and FEDER Funds, Ref.~CLU-2023-1-05.

\bibliographystyle{apsrev4-1}
\bibliography{References2}

\end{document}